\documentclass[prb,twocolumn,groupedaddress,showpacs,noeprint,nofootinbib,superscriptaddress]{revtex4-1}

\usepackage{epsfig,amsmath,amssymb,graphics,calc}
\usepackage[dvipsnames]{xcolor}
\usepackage[utf8]{inputenc}
\usepackage[english]{babel}
\usepackage[colorlinks,linkcolor=purple,citecolor=green!50!black,urlcolor=blue!70!black]{hyperref}
\usepackage{braket}
\usepackage{mathtools}

\DeclareMathOperator{\e}{e}
\newcommand{\di}{\mathrm d}
\newcommand{\vol}{\mathcal{V}}
\newcommand{\recK}{r}
\newcommand{\eerg}{\varepsilon_{\rm erg}}
\newcommand{\LPENS}{Laboratoire de Physique de l'Ecole normale sup\'erieure, ENS, Universit\'e PSL, CNRS, Sorbonne Universit\'e, Universit\'e de Paris, F-75005 Paris, France}
\newcommand{\KCLMaths}{Department of Mathematics, King's College London, Strand, London, WC2R 2LS,
	United Kingdom}

\begin{document}

\title{Out of equilibrium Phase Diagram of the Quantum Random Energy Model}

\author{Giulio Biroli}
\affiliation{\LPENS}
\author{Davide Facoetti}
\affiliation{\LPENS}
\affiliation{\KCLMaths}
\author{Marco Schir\'o}
\affiliation{JEIP, USR 3573 CNRS, Coll\`ege de France, PSL Research University, F-75321 Paris, France}
\author{Marco Tarzia}
\affiliation{LPTMC, CNRS-UMR 7600, Sorbonne Universit\'e, 4 Pl. Jussieu, F-75005 Paris, France}
\affiliation{Institut  Universitaire  de  France,  1  rue  Descartes,  75231  Paris  Cedex  05,  France}
\author{Pierpaolo Vivo}
\affiliation{\KCLMaths}

\begin{abstract}
In this paper we study the out-of-equilibrium phase diagram of the quantum  version  of  Derrida's  Random  Energy Model, which is the simplest model of mean-field spin glasses. We interpret its corresponding quantum dynamics in Fock space as a one-particle problem in very high dimension to which we apply different theoretical methods tailored for high-dimensional lattices: the Forward-Scattering Approximation, a mapping to the Rosenzweig-Porter model, and the cavity method. 
Our results indicate the existence of two transition lines and three distinct dynamical phases: a completely many-body localized phase at low energy, a fully ergodic phase at high energy, and a multifractal ``bad metal'' phase at intermediate energy. In the latter, eigenfunctions occupy a diverging volume, yet an exponentially vanishing fraction of the total Hilbert space. We discuss the limitations of our approximations and the relationship with previous studies.
\end{abstract}


\maketitle

\section{Introduction}
As discovered over $10$ years ago by the seminal work of Basko, Aleiner, and Altshuler,~\cite{BaskoAleinerAltshuler} isolated disordered interacting many-body systems can show absence of transport and thermalization even at finite energy density if the disorder is strong enough.
This is known as Many-Body Localization (MBL) and is a purely quantum phenomenon which occurs due to Anderson localization in the Fock space as the result of the interplay of disorder, quantum fluctuations, and interactions,\cite{BaskoAleinerAltshuler,Altshuler1997,Gornyi2005} and gives rise to a completely new mechanism for ergodicity breaking, that produces a robust dynamical phase of matter which is stable within a range of interaction and other Hamiltonian parameters.
This remarkable phenomenon has attracted considerable interest recently---see Refs.~[\onlinecite{Altman2015Review,Nandkishore2015,AbaninPapic2017,AletLaflorencie2018,Abanin2019RMP}] for recent reviews---as it implies that the long-time properties of MBL systems cannot be described by the conventional ensembles of quantum statistical mechanics: They can remember, forever and locally, information about their initial conditions.

Although significant and exciting progress has been made in
understanding these phenomena in recent years, both in theory\cite{Altman2015Review,Nandkishore2015,AbaninPapic2017,AletLaflorencie2018,Abanin2019RMP} and experiment,\cite{Schreiber2015,Bordia2016,Choi2016} there still remain many open issues.
A first set of open questions is about the nature (the universality class) of the MBL phase
transition between the thermal and localized phases as the randomness is increased. 
This transition is an eigenstate phase transition, marked by a sharp change in properties of the many-body wave-functions and thus in the dynamics of the system. However, the behavior of many-body eigenstates in the Hilbert space is
not firmly established:\cite{Luitz2015,LuitzBarLev2017,Mace2018} For instance, it is still debated whether there is only one phase transition, or could there possibly be some sort of intermediate phase that is neither fully localized nor fully thermal, where eigenstates are delocalized but non-ergodic,\cite{Altshuler1997,Pino2016,Pino2017,TorresHerrera2017} called the ``bad metal'' phase.\cite{BaskoAleinerAltshuler}
Investigations of the MBL transition so far have mostly been numerical studies based on Exact Diagonalization (ED) of relatively small one-dimensional systems,\cite{OganesyanHuse2007,PalHuse2010,Luitz2015} and how to do a proper finite-size scaling analysis of these numerical data remains unclear.\cite{Suntajs2019,Abanin2019}
Also in experiments it is challenging to access the very long time, and possibly also long length scales, on which the critical behavior develops.
Another frontier of the field is directed towards understanding the existence of MBL in higher dimensions,\cite{DeRoeck2017,Luitz2017,Ponte2017} and its relationship with other form of ergodicity breaking such as quantum glassiness.\cite{Laumann2014,Baldwin2016,Baldwin2017,Faoro2019}

In this context, exactly solvable, mean-field-like, and simplified toy models might naturally play an important role in making some progress, at least partially, in these directions and in improving our understanding of MBL. 
Also developing new techniques and tools to tackle analytically or semi-analytically the transition and its properties might provide an important step forward to shed new lights on some of the problems mentioned above.
With this in mind, in this paper we study the out-of-equilibrium phase diagram of the quantum  version of  Derrida's  Random  Energy Model,\cite{REM} which is the simplest toy model of mean-field spin glasses. 
The quantum model's equilibrium phase diagram has been studied before\cite{Goldschmidt1990,Jorg2008} and a glassy phase was  identified  at  low  temperature  and  small  transverse magnetic field. The MBL transition of the QREM was also investigated previously.\cite{Laumann2014,Baldwin2016,Baldwin2018,Faoro2019,Parolini2020,Smelyanskiy2019} In Refs.~[\onlinecite{Laumann2014}] and~[\onlinecite{Baldwin2016}] the presence of a mobility edge separating ergodic eigenstates  from many-body localized ones was established, based on EDs of small systems and on a perturbative expansion
built on the Forward-Scattering Approximation (FSA).\cite{anderson} 
A later study identified three distinct dynamical phases, referred to as trapped, tunneling and excited phases in the context of quantum optimization problems.\cite{Baldwin2018} The interpretation of those phases, and in particular of the phase right above the MBL localized phase, has been recently put into question. In particular, in Ref.~[\onlinecite{Faoro2019}] the MBL phase has been identified with an hyperglass where dynamics is practically absent while the entire phase above $T_{\rm MBL}$ with a dynamical glass phase (or bad metal) characterized by non-ergodic extended (NEE) eigenstates. 
In Ref.~[\onlinecite{Smelyanskiy2019}] the authors derive an estimate of the transition to NEE eigenstates in agreement with Ref.~[\onlinecite{Faoro2019}], and argue that the NEE phase is layered in an alternating sequence of two distinct subphases.
The dynamical population transfer protocol on the QREM was further analyzed in Ref.~[\onlinecite{Parolini2020}], yielding a numerical estimation of the dynamical phase diagram and of the fractal dimensions of the eigenstates in the NEE regime.

In this work we revisit the problem of the out of equilibrium phase diagram of the QREM using two complementary techniques, the first based on the FSA and on a mapping onto a paradigmatic random matrix model, the Rosenzweig-Porter (RP) model,\cite{Kravtsov2015,Facoetti2016} and the second  based on a generalization of the self-consistent theory of localization\cite{ATA,Logan2019,RoyLogan2020,RoyLogan2020-2} (hereafter called the ``cavity approach'') designed to take into account the local structure of the Hilbert space of the QREM.   In agreement with Ref.~[\onlinecite{Faoro2019}] we find a multifractal ``bad metal'' phase in a broad range of intermediate energies, where eigenfunctions are delocalized but non-ergodic, and out-of-equilibrium relaxation to thermal equilibrium is expected to be very slow\cite{Faoro2019,Parolini2020} (exponential in the system size). We also obtain a second transition into a fully delocalized ergodic phase at higher temperatures.

The paper is organized as following: In section~\ref{sec:model} we introduce the QREM model	 and recall basic properties of its equilibrium phase diagram. Section~\ref{sec:HS} describes the mapping to a single particle tight-binding Anderson problem in Hilbert space. Section~\ref{sec:RP} provides qualitative arguments for the phase diagram within the FSA. Section~\ref{sec:cavity} contains the cavity approach and the numerical results found with this method.
In Section~\ref{sec:PD} we summarize the results found with our approximations and discuss their relationship with previous results.
Section~\ref{sec:toy} put forward an interpretation of the results based on a family of auxiliary Anderson ``toy'' models. Finally, in Sec.~\ref{sec:conclusions} we provide concluding remarks and perspective for future studies. Several technical details are reported in the Appendices~\ref{app:clusters}-\ref{app:analytic}.

\section{The model} \label{sec:model}
The  Quantum  Random Energy  Model  (QREM)  for $N$ spin-$1/2$s is defined by the following Hamiltonian:
\begin{equation}
\label{eq:HQREM}
{\cal H}_{\rm QREM} = E(\{ \hat{\sigma}_a^z \}) - \Gamma \sum_a \hat{\sigma}_a^x \, ,
\end{equation}
where $\Gamma$ is the transverse field, and $E(\{ \hat{\sigma}_a^z \})$ is a random
operator diagonal in the $\{ \hat{\sigma}_a^z \}$ basis, 
which takes $2^N$ different values for the $2^N$ configurations of the $N$ spins in the $z$-basis,
identically and independently distributed according to:
\begin{equation} \label{eq:PE}
P(E) = \frac{e^{-E^2/N}}{\sqrt{\pi N}} \, .
\end{equation}
Such natural choice of the scaling of the random many-body energies insures that they are with high
probability contained in the interval $[-N \sqrt{\ln 2}, +N \sqrt{\ln 2}]$ in the thermodynamic limit.
Throughout the paper we will denote by $\varepsilon = E/N$ the intensive energy per spin corresponding to the extensive energy $E$.
A concrete implementation of the $E(\{ \hat{\sigma}_a^z \})$ is given by the $p\to \infty$ limit of the fully-connected $p$-spin model: $E(\{ \hat{\sigma}_a^z \}) = 
\lim_{p \to \infty} \sum_{a_1, \ldots, a_p} J_{a_1,\ldots,a_p} \hat{\sigma}_{a_1}^z \cdots \hat{\sigma}_{a_p}^z$,
where the $J_{a_1,\ldots,a_p}$ are i.i.d.~Gaussian random variables. In consequence, the diagonal part of the Hamiltonian corresponds to a mean-field spin-glass model, and exhibits all the fundamental features of the so-called ``random first-order theory'',\cite{rfot} with a 1-step replica symmetry breaking glass transition.\cite{REM}

The equilibrium properties (in the canonical ensemble) of the QREM are well established.\cite{Goldschmidt1990,Jorg2008} 
At low transverse field, it displays the same transition as the classical model between the paramagnetic and the glass phase
at the Kauzmann temperature $T_K = 1/(2 \sqrt{\ln 2})$. All thermodynamic quantities are identical to the
$\Gamma=0$ classical REM. For $T<T_K$ the system freezes in its ground state at $\varepsilon_{\rm GS} = - \sqrt{\ln 2}$.
The cassical model has also a dynamical ergodicity-breaking transition below which the time to reach thermal equilibrium is exponentially large in the system size \cite{baity2018activated}. However, differently from the $p$-spin models with finite $p$, for which the dynamical transition temperature $T_d$ is finite, in the $p \to \infty$ limit $T_d \to \infty$, due to the fact that the random energies and spin configurations are totally uncorrelated and flipping a single spin can change the energy by a large amount.
Within the semiclassical approximation of Ref.~[\onlinecite{Goldschmidt1990}] $T_d$ stays infinite even when quantum fluctuations are turned on ($\Gamma > 0$).

At large magnetic field, $\Gamma>\Gamma_c(T)$ the system is a standard quantum paramagnet, and the  REM  term  in the Hamiltonian does not influence the equilibrium physics of this phase. The first-order transition between these two regions takes place at $\Gamma_c(T)$, which is equal to $\sqrt{\ 2} \approx 0.833$ for $T \to 0$ and to $\sqrt{2}/2 \approx 0.707$ for $T \to \infty$. In this paper we will only focus on the small-$\Gamma$ region of the phase diagram (i.e., $\Gamma < \sqrt{2}/2$).

\section{Mapping to Anderson Localization on the Hypercube} \label{sec:HS}

The QREM defined by Eq.~(\ref{eq:HQREM}) can be viewed as the simplest many-body model that displays Anderson localization in
its Hilbert's space: If one chooses as a basis the tensor product of the simultaneous eigenstates of the operators $\sigma_a^z$,
the Hilbert space of the many-body Hamiltonian is a $N$-dimensional hypercube of $\vol = 2^N$ sites.
One can map a configuration of $N$ spins to a corner  
of the $N$-dimensional hypercube  by  considering
$\sigma_a^z = \pm1$ as  the  top/bottom  face  of  the  cube's $a$-th  dimension. 
The random part of the Hamiltonian is by definition diagonal on this basis, and gives {\it uncorrelated} 
random energies on each site orbital of the hypercube: 
At $\Gamma = 0$ the many-body eigenstates of Eq.~(\ref{eq:HQREM}) are simply product states of the form $\vert \sigma_1^z \rangle \otimes \vert \sigma_2^z \rangle \otimes \cdots \otimes \vert \sigma_N^z \rangle$, and the system is fully localized.
The interacting part of the Hamiltonian acts as single spin flips on the configurations $\{ \sigma_a^z \}$, and plays the role the hopping rates connecting ``neighboring'' sites in the configuration space.
The many-body quantum dynamics
is then recast as a single-particle non-interacting tight-binding Anderson model for spinless electrons in a disordered potential living  on the $2^N$ corners of an hypercube in $N$ dimensions, with the spin configurations being ``lattice sites'', $\vert i \rangle \equiv \vert \{ \sigma_a^z \} \rangle$,  and the transverse field playing the role of the hopping amplitude between neighboring sites:
\begin{equation} \label{eq:H}
{\cal H} = - \Gamma \sum_{\langle i,j \rangle} \left( \vert i \rangle \langle j \vert 
+ \vert j \rangle \langle i \vert 
\right ) + \sum_{i=1}^\vol E_i \, \vert i \rangle \langle i \vert 
\, ,
\end{equation}
where $\langle i,j \rangle$ denotes nearest-neighbors on the hypercube,
$\vol = 2^N$ is the total number of sites,  
and $\Gamma$ is the hopping kinetic energy scale. 
This mapping is {\it exact},
in the sense that the Hamiltonians~(\ref{eq:HQREM}) and~(\ref{eq:H}) have {\it the same} eigenvalues and the eigenvectors (when the simultaneous eigenstates of the operators $\sigma_i^z$ is chosen as a basis).
However, for a generic interacting many-body Hamiltonian in finite dimensions the random energies defined on neighboring corners of the hypercube are strongly correlated, as they correspond to many-body configurations which only differ by a single spin flip, while for the QREM 
$E_i$ are i.i.d.~random variables distributed according to Eq.~(\ref{eq:PE}), since its distinguishing feature is precisely the absence of such correlations.

\section{Estimate of the  Out of equilibrium phase diagram within the Forward-Scattering approximation} \label{sec:RP}

As discussed in Section~\ref{sec:HS}, the QREM can be mapped to an Anderson model on a hypercube with $\vol=2^N$ sites, labeled by $\sigma^z$ configurations.
The typical tunneling rate between two configurations depends on their energy and on the Hamming
distance $Nx$ between them (the Hamming distance $Nx$ is defined as the minimum number of spin flips which separate the two configurations, with $0<x<1$).
Since the energy levels are independent, the typical number of configurations of energy $|E|=N \varepsilon$ and at distance $Nx$ from a given configuration is
\begin{equation}
  \mathcal{N}_\varepsilon(x) = \binom{N}{N x} \frac{\e^{-N \varepsilon^2}}{\sqrt{\pi N}} \, .
\end{equation}
Here
we estimate the matrix elements $\mathcal{M}(\varepsilon,x)$ between these two configurations by perturbation theory in $\Gamma$, using the FSA.\cite{Laumann2014,Pietracaprina2015,Tarzia2020}
This consists in assuming that the matrix element between two configurations at distance $x$ is given by the product of the matrix elements obtained along the $Nx$ spin flips that connect the two configurations,
ignoring ``loopy'' contributions in which spins are flipped twice since they contribute at higher order in perturbation theory. Almost all states have energy $\mathcal{O}(\sqrt{N})$, while $E\approx\mathcal{O}(N)$, therefore we take the energy differences appearing in the denominators of the preturbative expansion to be $E$. Since there are $(N x)!$ such contributions, corresponding to the $(N x)!$ to connect two configurations $Nx$ spin flips away, the resulting  matrix element reads
\begin{equation} \label{eq:Mex}
\mathcal{M}(\varepsilon,x) \approx \left(\frac{\Gamma}{N \varepsilon}\right)^{Nx} \left(Nx\right)! \, .
\end{equation}
In the case of non-interacting particles in a disordered potential, the Mott argument for hybridization states that the metal-insulator transition occurs when the number of sites in resonance with a given site $i$ stays finite in the thermodynamic limit.\cite{Bogomolny2018,Nosov2019,Kravtsov2020}
Based on the analogy with single particle Anderson localization
one can thus characterize the localized phase by the condition (Mott criterion):
\begin{equation} \label{eq:AL}
 \lim_{N\to\infty}\mathcal{N}_\varepsilon(x)  |\mathcal{M}(\varepsilon,x)| = 0 \, ,
 \end{equation}
and from this estimate the point at which Many-Body localization of the QREM takes place.

Analogously, the ergodicity breaking transition can be estimated from the Fermi Golden Rule.\cite{Bogomolny2018,Nosov2019,Kravtsov2020,Faoro2019} In fact the spreading amplitude $\mathcal{N}_\varepsilon(x)  |\mathcal{M}(\varepsilon,x) |^2$ 
quantifies the escape rate of a particle created at a given site $i$.
When this amplitude is much smaller than the spreading   of energy levels  due to disorder, 
 then ergodicity is broken:
 \begin{equation} \label{eq:ergo}
 \lim_{N\to\infty}\mathcal{N}_\varepsilon(x)  |\mathcal{M}(\varepsilon,x) |^2 = 0 \, .
 \end{equation}
The non-ergodic extended phase is thus realised when
\begin{equation}
    \mathcal{N}_\varepsilon(x)  |\mathcal{M}(\varepsilon,x) |^2 \to 0 {\rm~~~and~~~}
\mathcal{N}_\varepsilon(x)  |\mathcal{M}(\varepsilon,x) | \to \infty \, .
\end{equation}
This means that although a given state is in resonances with many other states of energy $\varepsilon$ and at distance $Nx$ from it, the number of those resonances is not enough for the quantum dynamics to decorrelate \textit{in a finite time} from the initial condition.

These two criteria for localization and ergodicity breaking can be illustrated using the Rosenzweig-Porter (RP) model\cite{Kravtsov2015,Facoetti2016,DeTomasi2018,Bogomolny2018} as a benchmark.
For the sake of clarity, here we briefly recall the definition of the RP model, whose Hamiltonian is a matrix of size $\vol \times \vol$ given by the sum of two terms,
\begin{equation}
  H_{RP} = {\cal E} + \frac{\mu}{\vol\strut^{\gamma/2}} {\cal G} \ ,
\end{equation}
where $\mathcal{E}_{ij} = E_i \delta_{ij}$ is diagonal with i.i.d.~entries (the distribution does not matter as long as it has finite variance), $\mu$ is a constant of order one (whose  value is unimportant), and ${\cal G}$ is a random matrix drawn from the Gaussian Unitary Ensemble (GUE) with unit variance.
The latter mimics an ergodic system (e.g.~the clean lattice), while ${\cal E}$ represents the on-site disorder.
The parameter $\gamma$ acts in the RP model as a proxy of the disorder strength: at large $\gamma>2$, the GUE contribution is suppressed, and the systems is localized; at small $\gamma<1$, the system is ergodic, while the regime $1<\gamma<2$ is special, with delocalized but nonergodic wavefunctions which typically occupy $\vol^{2-\gamma}$ sites close in energy. The criteria for these two transitions are exactly the ones we introduced above for Anderson localization~(\ref{eq:AL}) and ergodicity breaking~(\ref{eq:ergo}).

Specializing now to the $\mathcal{N}_\varepsilon(x)$ levels at energy $\varepsilon$ and Hamming distance $Nx$ between them of the QREM, we can map the effective (typical) transition rates~(\ref{eq:Mex}) onto 
an effective $\mathcal{N}_\varepsilon(x)  \times \mathcal{N}_\varepsilon(x)$ RP random matrix 
with off-diagonal matrix elements roughly scaling as $\mathcal{M}(\varepsilon,x) \propto \mathcal{N}_\varepsilon(x)^{-\gamma}$.
The effective exponent $\gamma$ associated to a given set of configurations $(\varepsilon,x)$ defined as $\gamma = - 2 \ln {\cal M} (\varepsilon,x) / \ln {\cal N}_\varepsilon (x)$. 
We therefore expect that the transition from Anderson localization to nonergodic extended states for the QREM  happens when 
at least one $x$-sector 
becomes delocalized, i.e.
\begin{eqnarray}
\max_{x}\mathcal{N}(\varepsilon,x)\mathcal{M}(\varepsilon,x)\approx \max_x\e^{N f_1(x, \varepsilon,\gamma)} \gtrsim 1 ,
\end{eqnarray}
where, using Stirling's approximation for the factorials,
\begin{equation}\label{eq:rp-f1}
f_1(x,\varepsilon,\Gamma)=x\ln\left(\frac{\Gamma}{e \varepsilon}\right)
-(1-x)\ln(1-x)-\varepsilon^2\ .
\end{equation}
If $\Gamma<\varepsilon$, $f_1$ is always negative. Otherwise, it has a non-negative maximum at $x_1^* = 1 - \varepsilon/\Gamma$,
which determines the mobility edge $\Gamma_{\rm MBL}(\varepsilon)$ through the implicit equation
\begin{equation}\label{eq:qp-dyn-f1}
f_1(x_1^*,\varepsilon,\Gamma_{\rm MBL}) = \frac{\varepsilon}{\Gamma_{\rm MBL}} -\varepsilon^2 +\ln\frac{\Gamma_{\rm MBL}}{\e \varepsilon}=0 .
\end{equation}
This is the same result obtained through a similar argument in Ref.~[\onlinecite{Laumann2014}].

Similarly, full ergodicity is recovered  by requiring that at least one 
$x$-sector becomes ergodic,
\begin{eqnarray}\label{eq:rp-ergo}
\max_{x}\mathcal{N}(\varepsilon,x)|\mathcal{M}(\varepsilon,x)|^2\approx \max_x\e^{N f_2(x, \varepsilon,\gamma)} \gtrsim 1 ,
\end{eqnarray}
with
\begin{equation}\label{eq:rp-f2}
f_2(x,\varepsilon,\Gamma)=
x\ln x - (1-x) \ln(1-x)-\varepsilon^2+ 2 x\ln\left(\frac{\Gamma}{e \varepsilon}\right)\ .
\end{equation}
If $\Gamma<2\varepsilon$, $f_2$ is always negative. Otherwise, it has a non-negative maximum at $x_2^*=1/2(1 + \sqrt{1 - 4 (\varepsilon/\Gamma)^2})$, which gives a different implicit equation for the ergodic transition $\Gamma_{\rm erg}(\varepsilon)$.

\begin{figure}
	\includegraphics{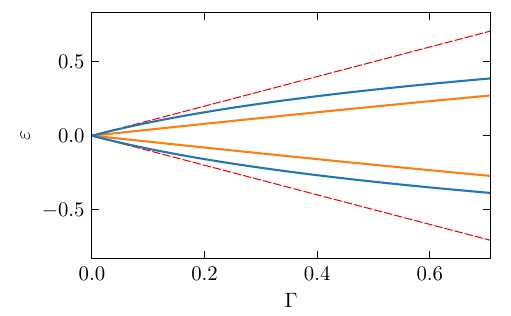}
	\caption{Localization (blue) and ergodicity (orange) transition lines,
	obtained from the mapping to the RP model, Eqs.~(\ref{eq:rp-f1},\ref{eq:rp-f2}).
The limits of the $y$ axis coincide with the edges of the many-body spectrum ($|\varepsilon| < \sqrt{\ln2}$). The red dashed lines correspond to $\varepsilon= \pm \Gamma$ (see Refs.~[\onlinecite{Laumann2014}] and~[\onlinecite{Baldwin2016}]).}
    \label{fig:RP-micro-phase}
\end{figure}

\begin{figure}
	\includegraphics{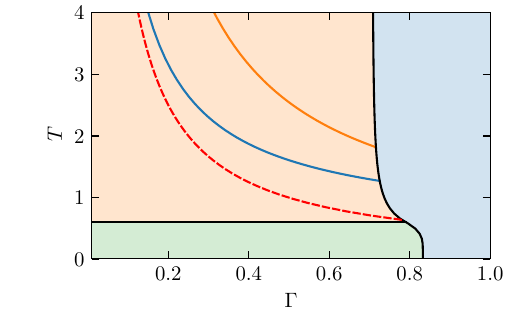}
	\caption{Ergodicity and localization transition lines (Fig.~\ref{fig:RP-micro-phase}), transposed on the canonical phase diagram, $T=1/2\varepsilon$.}
	\label{fig:RP-canonical}
\end{figure}

Expanding the solutions to the two implicit equations around $\varepsilon=0$, we find $\Gamma_{\rm MBL}\approx\varepsilon$,\cite{Laumann2014,Baldwin2016,Baldwin2018,Faoro2019}, and $\Gamma_{\rm erg}\approx  \varepsilon/2$. 
The analogy with the RP model therefore indicates that the QREM also undergoes two separate localization and ergodicity transitions.
The estimates for the transition lines obtained in this way are shown in Figs.~\ref{fig:RP-micro-phase} and \ref{fig:RP-canonical}.

Note that the phase diagram of Fig.~\ref{fig:RP-micro-phase} implies that the non-trivial localized and non-ergodic behaviours are found at non-zero energy density. Hence the 
fractions of localized and NEE states is exponentially small in $N$ and vanish as $N \to \infty$, while the vast majority of states around zero energy density are fully ergodic, in agreement with Ref.~[\onlinecite{RoyLogan2020}].

Qualitatively, these results are in  agreement with the recent analysis of Refs.~[\onlinecite{Faoro2019}] and~[\onlinecite{Smelyanskiy2019}], which are based on a different (and more thorough) strategy to estimate the off-diagonal tunneling rates between different spin configurations in the Hilbert space, and are also in agreement with the numerical estimations of Ref.~[\onlinecite{Parolini2020}]. However, while Eq.~(\ref{eq:rp-f2}) predicts that the transition line from the fully ergodic to the dynamical glass (bad metal) regime is at finite energy density $\eerg$, within the approach of Refs.~[\onlinecite{Faoro2019}] and~[\onlinecite{Smelyanskiy2019}] the transition line is squeezed to zero energy density, i.e., infinite temperature. 
We will come back to this point in Sec.~\ref{sec:PD}.

\section{Cavity approach}
\label{sec:cavity}

\subsection{Cluster approximation on the hypercube}

In large spatial dimensions the neighbors of a given site are organized in a hierarchical way (i.e., the fraction of short loops is suppressed) and their number grows exponentially with the distance. These are distinctive features of tree-like structures. 
In fact, it was argued originally in~[\onlinecite{Altshuler1997}] that the (non-interacting) Anderson model on the Bethe lattice,
first introduced and studied in Ref.~[\onlinecite{ATA}], can be thought as a toy model for MBL (see also Refs.~[\onlinecite{Gornyi2005,Jacquod1997,LoganWolynes1987,LoganWolynes1990,Bigwood1998}] for a similar analysis and Ref.~[\onlinecite{DeLuca2013}] for a quantitative investigation of these ideas).

Since on tree-like structures the model~(\ref{eq:H}) allows, in principle, for an exact solution,\cite{ATA} which yield 
the diagonal elements of the resolvent matrix, 
assuming that for large enough $N$ the hypercube is well approximated by a Bethe lattice 
provides a very simple way to investigate analytically, although approximately, the spectral properties of the eigenvectors of the QREM (see Refs.~[\onlinecite{Logan2019,RoyLogan2020,RoyLogan2020-2}] and~[\onlinecite{RoyLazarides2020}] for similar approaches in the context of MBL).

The simplest approximation consists in taking 
Random-Regular Graphs (RRGs) of $\vol = 2^N$ sites and fixed connectivity $N$ as the underlying lattices mimicking the Hilbert space,\cite{DeLuca2013} i.e., random lattices which have locally a tree-like structure but have loops whose typical length scales as $\ln \vol \propto N$ and no boundary, and which are statistically translationally invariant.\footnote{The properties of random-regular graphs have been extensively studied. For a review see~\cite{WormaldRRG}}
In practice, this corresponds to shuffling the position of the sites and/or rewiring the connections of the hypercube in a random way, keeping the total number of sites and the local connectivity of the lattice fixed.

There is however a 
potentially important issue related to the Bethe approximation, which we explain below.
The spectrum of the kinetic term of~(\ref{eq:H}) is given by the Density of States (DoS) of the adjacency matrix of the $N$-dimensional hypercube, which coincides with the distribution of the eigenvalues of the second term of the Hamiltonian~(\ref{eq:HQREM}), i.e.,  a simple paramagnet, with energies contained in the interval $E \in [-N \Gamma, N \Gamma]$:
 \begin{equation} \label{eq:rhoIHC}
 \rho_I^{\rm HC} (E) = \frac{\Omega(E)}{\Gamma 2^{N+1}} \, ,
 \end{equation}
 where the term $\Gamma 2^{N-1}$ in the denominator is a normalization factor that insures that $\int_{-N \Gamma}^{+N \Gamma}  \rho_I^{\rm HC} (E) \, {\rm d} E = 1$, and
 $\Omega(E)$ is the number of spin configurations at energy $E$:
 \begin{equation}
 \Omega(E) = \binom{N}{(N+E/\Gamma)/2} \sim \Omega_0  e^{N s(\varepsilon/\Gamma)} \, .
 \end{equation}
 $\Omega_0$ is a normalization factor and $s(\varepsilon/\Gamma)$ is the entropy per spin at large $N$ (apart from logarithmic corrections) for a polarization $m = \varepsilon / \Gamma = \langle \sigma^x \rangle$ of the spins in the $x$ direction:
 \begin{equation}
 s ( m ) = \ln(2) - \frac{1 + m}{2} \ln  ( 1 + m ) - \frac{1 - m}{2} \ln ( 1 - m ) \, .
 \end{equation}
 Approximating the hypercube as a tree-like lattice amounts to replacing Eq.~(\ref{eq:rhoIHC})
 with 
 the spectrum of the adjacency matrix of a Bethe lattice of connectivity $N$ which, for large enough $N$ tends asymptotically to a semicircle law:
 \begin{equation} \label{eq:rhoIRRG}
 \rho_I^{\rm RRG} (E) \approx
 \frac{\sqrt{4 \Gamma^2 N - E^2}}{8 \pi \Gamma^2 N} \, ,
 \end{equation}
 with support in the interval $E \in [-2 \Gamma \sqrt{N}, 2 \Gamma \sqrt{N}]$.
 For energies of order $O (\sqrt{N})$, where the vast majority of the eigenvalues is, $\rho_I^{\rm RRG}$ provides in fact a reasonably good estimation for the true DoS, $ \rho_I^{\rm HC}$ (see fig.~\ref{fig:rhoI} of Appendix~\ref{app:clusters} for a quantitative comparison).
 Yet, for {\it extensive} energies of $O(N)$ Eq.~(\ref{eq:rhoIRRG}) 
 completely neglects the exponentially small fraction of eigenvalues in the tails of the DoS, corresponding to strongly polarized spins in the $x$ direction.
Since at finite energy density $\varepsilon$ and $N$ sufficiently large, 
the energy $N \varepsilon$ will fall outside the edges of the semicircle~(\ref{eq:rhoIRRG}) which describes the spectrum of the delocalizing kinetic term within the Bethe approximation, 
the Anderson localization of the eigenfunctions of the Hamiltonian~(\ref{eq:H}) will occur in the far Lifshits tails of the DoS,\cite{BiroliSemerjianTarzia} and this might affect its properties. 
In other words, the system might appear as localized within the Bethe approximation due to the fact that some of the matrix elements associated to the kinetic term are artificially suppressed by approximating the hypercube as a RRG.

In order to overcome, at least partially, these limitations in this paper we put forward a cluster expansion, 
which takes into account, at least locally, the specific structure of the hypercube up to a certain distance (in particular, including all the shortest loops of length four, six, eight, etc.) and improves systematically the simplest, single-site, Bethe approximation. In practice, we consider clusters of $s=2^n$ neighboring corners on the hypercube (corresponding to spin configurations which differ by few spin flips only), 
and obtain self-consistent recursion equations for the $s \times s$ elements of the local resolvent matrix on each cluster by assuming that the clusters are on a tree-like structure.

\begin{figure}
\includegraphics[width=0.48\textwidth]{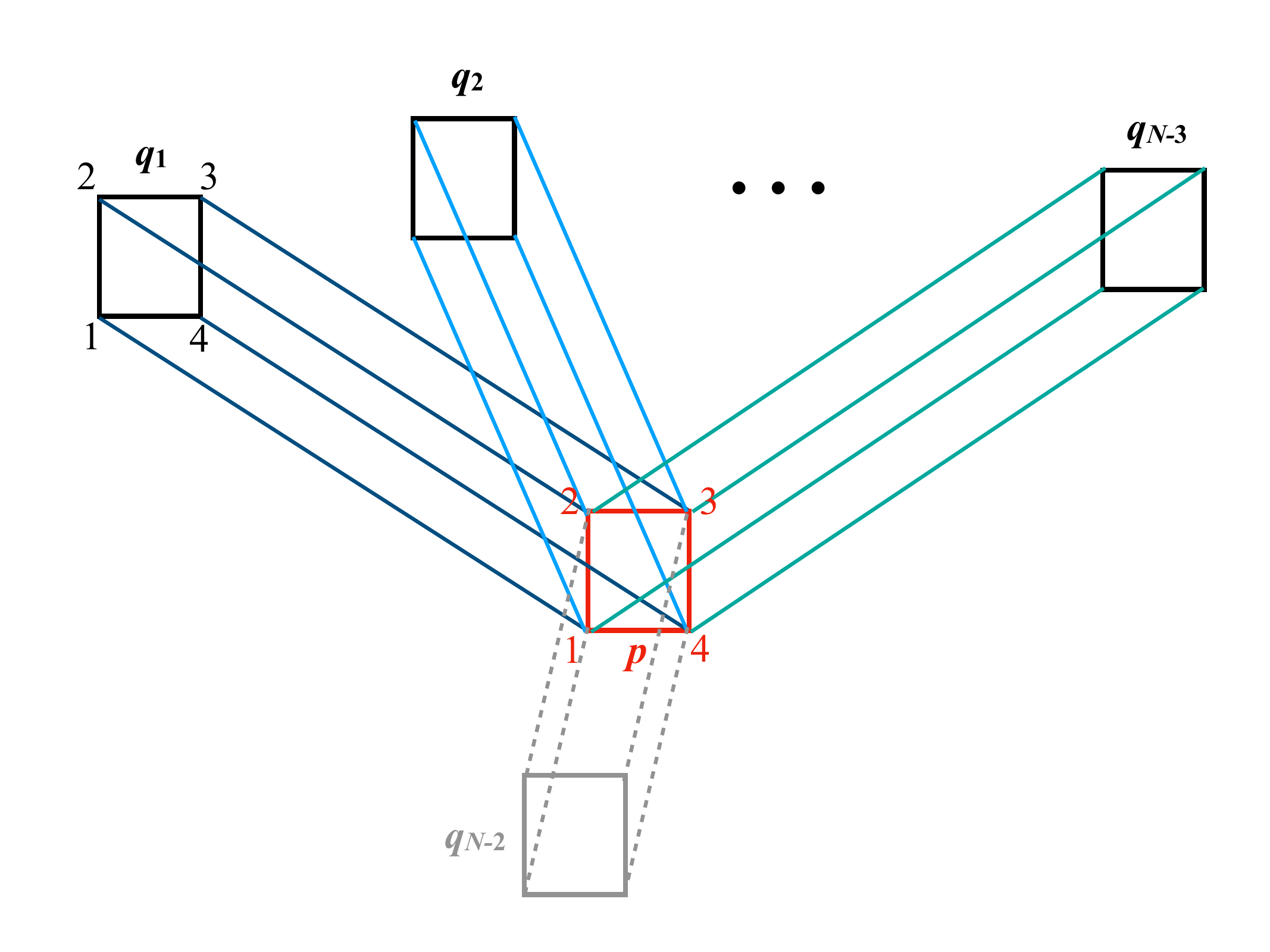}
\vspace{-0.2cm}
\caption {Schematic representation of the recursion step which yields the self-consistent equations for the $s \times s$ elements of the (cavity) resolvent matrix for clusters of 
$4$ sites. 
All loops up to length $12$ of the hypercube are treated exactly. }
\label{fig:cluster}
\end{figure}

 The standard single-site Bethe approximation corresponds to $n=0$, while the cases 
 $n=2$ is 
 schematically represented in fig.~\ref{fig:cluster} ($n=N$ corresponds, of course, to the exact solution of the problem). 
 We will take $n=2$ throughout.

For $n=2$ a plaquette of four corners corresponds to four spin configurations such as, e.g., 
\[
\begin{aligned}
\vert 1 \rangle &= \vert \uparrow , \uparrow , \sigma_3^z , \ldots , \sigma_N^z \rangle \, \\
\vert 2 \rangle &= \vert \uparrow , \downarrow , \sigma_3^z , \ldots , \sigma_N^z \rangle \, \\
\vert 3 \rangle &= \vert \downarrow , \downarrow , \sigma_3^z , \ldots , \sigma_N^z \rangle \, \\
\vert 4 \rangle &= \vert \downarrow , \uparrow , \sigma_3^z , \ldots , \sigma_N^z \rangle \, ,
\end{aligned}
\]
for any configuration $\{\sigma_a^z \}_{a=3,\ldots,N}$. The $N-2$ neighbors of such plaquette on the hypercube are found by flipping the $N-2$ spins $\sigma_3^z, \ldots, \sigma_N^z$ one by one. Two neighboring plaquettes of four sites are connected by four edges. The Hilbert space will be then approximated as a RRG of $2^N/4$ square plaquettes of connectivity $N-2$. More details are given in Appendix~\ref{app:clusters}.

One can show that within the cluster approximation the support of the spectrum of the kinetic term of the Hamiltonian~(\ref{eq:H}) becomes indeed broader and broader as $n$ is increased, 
$E_{\rm max} \approx \Gamma (2 \sqrt{N - n} + n)$ (see Appendix~\ref{app:clusters}).

One can easily obtain, at least formally, the self-consistent recursion relations for the elements of the 
resolvent matrix (or Green's functions)  
of the Hamiltonian~(\ref{eq:H}), defined as ${\cal G} (z) =  ({\cal H} - z {\cal I}  )^{-1}$, at any order $n$ of the cluster expansion.
The key objects are the so-called {\it cavity} resolvent matrices, 
$G_{p \to q} (z) = 
({\cal H}_{p \leftrightarrow q} - z {\cal I}  )^{-1}$, 
i.e., the 
resolvent matrices of modified Hamiltonians ${\cal H}_{p \leftrightarrow q}$ where all the $2^n$ edges between
the sites of the cluster $p$ and the sites of the cluster $q$ have been removed (gray dashed lines of fig.~\ref{fig:cluster}).
Let us assume, as explained above and as sketched in fig.~\ref{fig:cluster}, that at large enough $N$ the clusters occupy the vertices of a tree-like structure (at least locally), 
and let us imagine to take a given cluster $p$ and its neighbors $\{q_1, \ldots, q_{N-n} \}$. If one removes such cluster
from the graph, then the clusters $\{ q_1, \ldots, q_{N-n} \}$ are (quasi-)uncorrelated, since the lattice would break in $N-n$
semi-infinite (quasi-)disconnected branches (neglecting the large loops of length of order $N$). One then obtains (e.g., by Gaussian integration) the following iteration relations for the elements of the cavity resolvent matrix on the $s$ sites of the cluster:\cite{ATA,BiroliSemerjianTarzia,BiroliTarzia-Review}
\begin{equation} \label{eq:recursion}
\left[ G_{p \to q_k}^{-1} (z) \right]_{uv} = {\cal H}_{uv} 
- z \delta_{uv} -
\Gamma^2 \!\!\!\! \sum_{q_l \in \partial p / q_k}  \!\!\!\! [G_{q_l \to p} (z)]_{uv} \, ,
\end{equation}
where $z= N \varepsilon + i \eta$, $\eta$ is an infinitesimal imaginary regulator which regularize the
pole-like singularities in the right hand sides (see below), $\varepsilon$ is the intensive energy density around which one chooses to study the spectral properties, ${\cal H}_{uv}$ are the matrix elements of the Hamiltonian~(\ref{eq:H}) between the sites $u$ and $v$ belonging to the cluster $p$, and $\partial p/q$ denotes the set of all $N-n$ neighbors of the cluster $p$ except $q$. The indices $u,v=1, \ldots, s$ identifying the sites belonging to each clusters are chosen as in fig.~\ref{fig:cluster}. (Note that for each cluster with $N-n$ neighbors one can define $N-n$ cavity Green's functions and $N-n$ recursion relations of this kind.)

After finding the solution of Eqs.~\eqref{eq:recursion}, one can finally obtain the 
resolvent matrix of the original problem on a given cluster $p$ as a function of the
cavity Green's functions on the neighboring clusters:\cite{BiroliSemerjianTarzia,BiroliTarzia-Review}
\begin{equation} \label{eq:recursion_final}
\left[ {\cal G}_p^{-1} (z) \right]_{uv} = {\cal H}_{uv} - z \delta_{uv} - \Gamma^2 \!
\sum_{q_k \in \partial p}  \! \left [ G_{q_k \to p} (z) \right]_{uv} \, .
\end{equation}
For $n=0$ these equations simply give back the standard recursion relations for the Anderson model on the Bethe lattice (with connectivity $N$ and Gaussian iid random energies).\cite{ATA,BiroliSemerjianTarzia,BiroliTarzia-Review}
For $n=1$ and $n=2$ simple analytic expressions of the inverse of the local $s \times s$ resolvent matrices are known, which allows one to write simple recursion equations for its $s$ diagonal elements and its $s (s-1)/2$ off-diagonal elements (see Appendix~\ref{app:clusters} for more details).
For $n\ge3$, however, the local inversion involved in Eqs.~(\ref{eq:recursion}) and~(\ref{eq:recursion_final}) must be done numerically at each iteration step. 

There are essentially two ways, that we detail in Appendix~\ref{app:cavity}, to solve the recursion equations for the Green's function and obtain information on the spectral statistics at finite $N$. The most accurate strategy, to which we will refer to as ``Cluster Belief Propagation'' (C-BP) algorithm (see Ref.~[\onlinecite{BiroliTarzia-Review}] for a detailed explanation of this approach for the usual tight-binding Anderson model on the Bethe lattice), is to solve directly Eqs.~(\ref{eq:recursion}) and~(\ref{eq:recursion_final}) on random realizations of the hypercube of $2^N$ sites (i.e., $N$ spins) (see Appendix~\ref{app:cavity} for details). However, thanks to the fact that random energies $E_i$ of the  QREM are  uncorrelated, in order to access larger system sizes one can adopt another 
strategy, to which we will refer hereafter as ``Cluster Population Dynamics'' (C-PD) algorithm,\cite{MezardParisi2001PopDyn} which consists in interpreting the recursion relations for the Green's functions as equations for their probability distributions once the average over the disorder is taken. 
In fact, since $G_{p \to q}$ and ${\cal G}_{p}$ are random matrices, one can assume that averaging over the on-site random energies 
leads to 
functional equations on their probability distribution $Q (G)$ and ${\cal Q}({\cal G})$ (see Appendix~\ref{app:cavity} for details). 
Fig.~\ref{fig:PLIG} of App.~\ref{app:cavity} shows that the cluster approximation provides a quite accurate approximation of local observables, such as the distribution of the LDoS, as compared to exact diagonalization for small systems.

\subsection{Spectral statistics and the \texorpdfstring{$\eta \to 0^+$}{eta to zero} limit}

The statistics of the diagonal elements of the resolvent
gives---in the $\eta \to 0^+$ limit, see below---the spectral properties of $\mathcal{H}$.
In particular, the probability distribution of the Local Density of States (LDoS) at energy $E = N \varepsilon$ is given by:
\begin{equation} \label{eq:LDoS}
        \begin{aligned}
                \rho_i (\varepsilon) 
                & = \sum_\alpha | \langle i \vert \alpha \rangle |^2 \, \delta ( N \varepsilon - E_\alpha ) 
                =\lim_{\eta \to 0^+}  \frac{{\rm Im} {\cal G}_i (N \varepsilon + i \eta)}{\pi} \, ,
        \end{aligned}
\end{equation}
from which the average Density of States (DoS) is simply obtained as $\rho (\varepsilon)
 = (1/\vol) \sum_i \rho_i (\varepsilon) 
 = \lim_{\eta \to 0^+}  {\rm Tr} \, {\rm Im} {\cal G} (N \varepsilon + i \eta) / (\vol \pi)$.
 (We have defined ${\cal G}_i =  [{\cal G}_p]_{uu}$ with $i = 2^n p + u$, $p = 1, \ldots, 2^{N-n}$ and $u = 1, \ldots , s=2^n$).
Similarly, the spectral representation of the Inverse Participation Ratio of the eigenstates $\vert \alpha \rangle$ of energy $E_\alpha$ close to $N \varepsilon$ can be obtained as:
\begin{equation} \label{eq:IPR}
\begin{aligned}
        \Upsilon_2 (\varepsilon) &= \sum_i \vert \langle i \vert \alpha \rangle |^4 \, \delta ( N \varepsilon - E_\alpha ) \\
        & = \lim_{\eta \to 0^+}  
        \frac{\eta \sum_{i=1}^\vol |{\cal G}_{i} (N \varepsilon + i \eta) |^2}{\sum_{i=1}^\vol {\rm Im} {\cal G}_i (N \varepsilon + i \eta)} \, .
        \end{aligned}
\end{equation}
Another useful observables is the typical DoS:
\begin{equation} \label{eq:rhotyp}
\rho^{\rm typ} (\varepsilon) = \lim_{\eta \to 0^+} \frac{\exp \left( \vol^{-1} \sum_{i=1}^\vol \ln {\rm Im} {\cal G}_i (N \varepsilon + i \eta) \right)}
{\vol^{-1} \sum_{i=1}^\vol {\rm Im} {\cal G}_i (N \varepsilon + i \eta)} \, .
\end{equation}
However, at this point we encounter another difficulty which is due to the very unusual (and simultaneous) scaling with $N$ of the parameters of the Hamiltonian~(\ref{eq:H}). 
In fact, the dependence of the random energies and the connectivity of the lattice on $N$ produces a density of states that strongly concentrate around zero energy density, as naturally expected for many-body systems: At small $\Gamma$ one has that $\rho(E) \approx P(E) = e^{-E^2/N}/\sqrt{\pi N}$, while at large $\Gamma$ one expects that $\rho(E) \approx \Omega(E)/(\Gamma 2^{N+1})$, see Eq.~(\ref{eq:rhoIHC}).
Thus, in both cases the  mean  level spacing $\delta(E) = 1/(\vol \rho(E))$ is  well-defined  locally,  but  depends strongly (i.e., exponentially)  on  the
local energy density. In particular, for small $\Gamma$ one has that
\begin{equation} \label{eq:MLS}
\delta (\varepsilon) \approx \sqrt{\pi N} e^{N (\varepsilon^2 - \ln 2)} \, .
\end{equation}
In order for Eqs.~(\ref{eq:LDoS})-(\ref{eq:rhotyp}) to be well defined, the limit $\eta \to 0^+$ should be taken in such a way that the imaginary regulator goes to zero 
on the same scale as the mean level spacing.\cite{BiroliTarzia-Review}
Hence, studying the asymptotic behavior of the model at large $N$ implies varying simultaneously the following parameters:\cite{Logan2019}
\begin{itemize}
\item[-] The total number of sites diverges exponentially $\vol = 2^N$;
\item[-] The connectivity of the lattice grows as $N$; 
\item[-] The standard deviation of the random on-site energies grows as $\sqrt{N/2}$, according to 
Eq.~(\ref{eq:PE});
\item[-] The energy at which we study the system grows as $N \varepsilon$, with $\varepsilon$ of $O(1)$;
\item[-] The imaginary regulator vanishes exponentially as $\delta(\varepsilon) = 1/(\vol \rho(\varepsilon))$.
\end{itemize}
Thus, the $N\to \infty$ limit of the model~(\ref{eq:H}) is quite different from the usual thermodynamic limit of the standard (non-interacting) Anderson model (where one just takes the limit $\vol \to \infty$ and $\eta \to 0$ keeping fixed the other parameters) and might give rise to unusual scaling and new properties.

In the following we will be mostly interested in studying the dependence on $\Gamma$, $\varepsilon$, and $N$ of the probability distribution of the LDoS, Eq.~(\ref{eq:LDoS}), from which one can compute several spectral quantities of interest, such as, e.g.,  the IPR~(\ref{eq:IPR}) and the typical value of the LDoS~(\ref{eq:rhotyp}). 
In particular, in order to assess the ergodicity of the wave-functions, it is custom to introduce the fractal dimensions defined through the asymptotic behavior of these two latter quantities: 
\begin{equation} \label{eq:D1D2tilde}
\begin{aligned}
\Upsilon_2 (\varepsilon) & \sim \left [\tilde{\cal V}(\varepsilon)\right]^{- D_2 (\varepsilon)} \, ,\\
\rho^{\rm typ} (\varepsilon) & \sim \left [ \tilde{\cal V} (\varepsilon)\right]^{D_1 (\varepsilon) - 1} \, .
\end{aligned}
\end{equation}
This definition takes into account the actual volume of the portion of the phase space accessible at finite energy density $\varepsilon$, $\tilde{\cal V} (\varepsilon) = 2^N \rho (\varepsilon)$, since at finite energy density $\varepsilon$, ergodic eigenstates are uniformly spread over the hyper-surface at constant energy.
In the small-$\Gamma$ part of the phase diagram, as a first approximation one has that $\rho (E) \approx P(E)$.

\subsection{Numerical results for the fractal dimensions}

\begin{figure}
\includegraphics[width=0.49\textwidth]{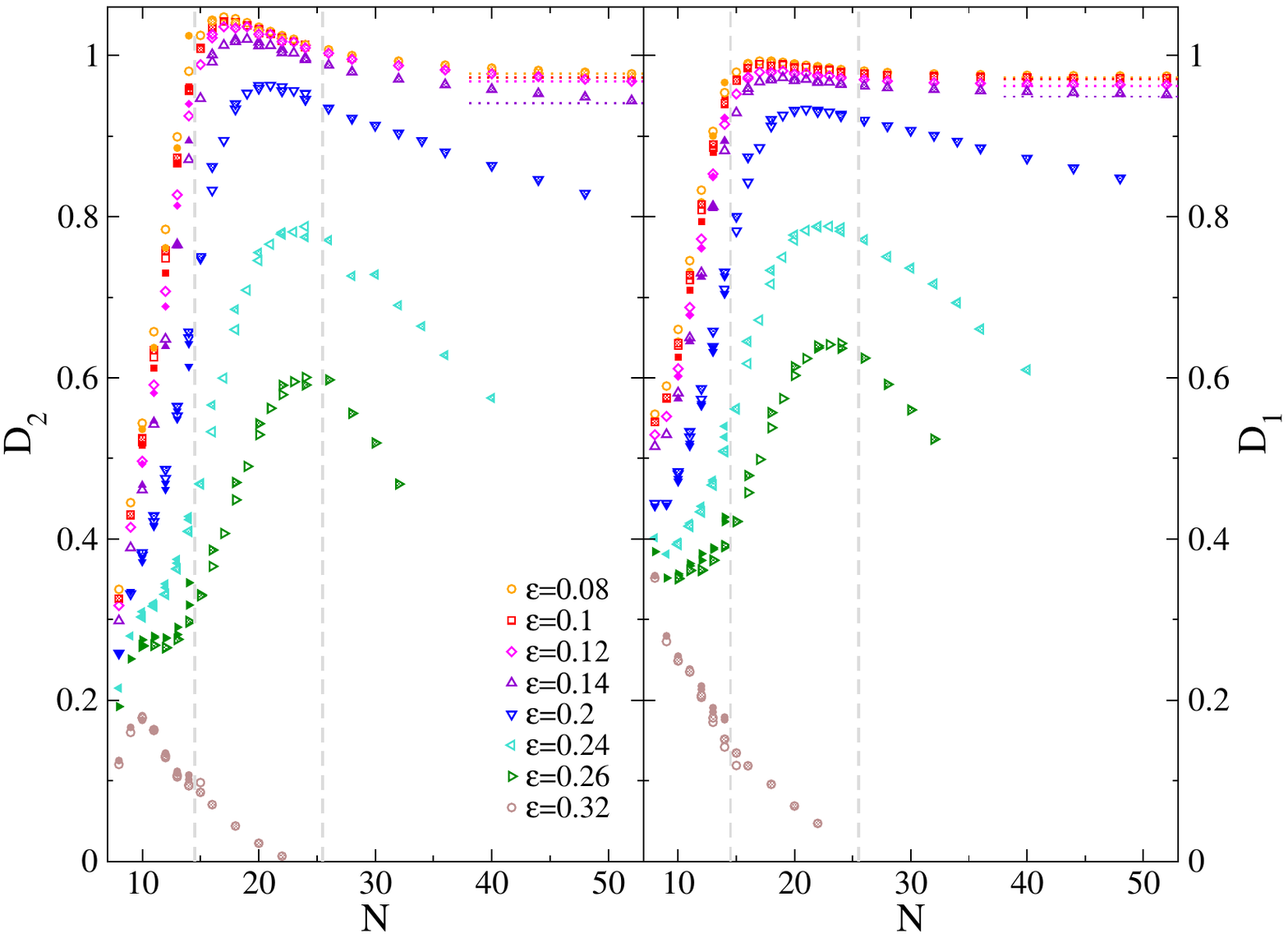}
\caption{\label{fig:D1D2} $D_2$ (left) and $D_1$ (right) versus $N$ for $\Gamma=0.2$ 
and several values of the energy density as indicated in the keys. Filled symbols correspond to the results obtained from ED, open symbols give the results of the C-BP approximation (with $n=2$), and shadow symbols correspond to the result of the C-PD approach (with $n=2$). Two independent ED data-sets, obtained with two different algorithms, are shown, to give the idea of the typical size of the errorbars. The vertical dashed lines represent the limits of the range of applicability of ED ($N \le 14$) and the C-BP approach ($N \le 25$). The horizontal dotted lines indicate the asymptotic values of $D_1$ and $D_2$ at large $N$ in the bad metal phase. (Note that the generalized fractal dimensions can become larger than $1$ for some intermediate values of $N$ and for $\varepsilon$ small enough, due to logarithmic corrections to the many-body DoS.)}
\end{figure}

In fig.~\ref{fig:D1D2}, where we plot $D_1$ (right panel) and $D_2$ (left panel) as a function of $N$ for $\Gamma=0.2$, 
and several values of the energy density $\varepsilon$, mostly  
on the delocalized side of the MBL transition and close to the mobility edge ($\varepsilon \lesssim \Gamma$). 
$D_1$ and $D_2$ are obtained as numerical derivatives of $\rho^{\rm typ}$ and $\Upsilon_2$ with respect to $\ln \tilde{\cal V}$.
The figure shows three data-sets, corresponding to the results obtained from EDs ($N \le 14$, filled symbols), the C-BP approach ($N \le 25$, $n=2$, open symbols), and the C-PD approximation ($N \le 50$, $n=2$, shadow symbols).

First of all, these plots demonstrate
that the C-BP and C-PD approximations are in reasonably good agreement with the exact results for all values of $\varepsilon$, at least in the range of system sizes accessible via EDs (see also Fig.~\ref{fig:PLIG} of App.~\ref{app:cavity} for a detailed comparison of the full probability distribution of the LDoS).

At moderate energy density both fractal dimensions show a clear non-monotonic dependence: $D_1$ and $D_2$ first rapidly increase with $N$, 
and then start to decrease slowly after going through a maximum. 
At small enough energy density, $\varepsilon \lesssim  0.14$, both $D_1$ and $D_2$ reach a  {\it finite} plateau strictly smaller than one at large $N$ (horizontal dotted lines). 
This behavior corresponds to genuine multifractal eigenstates, as recently predicted in Ref.~[\onlinecite{Faoro2019}], and is found in a broad range of energy density. 
The lower the energy, the higher are the plateau values reached by $D_1$ and $D_2$, i.e., the system gets closer and closer to full ergodicity as the energy density is decreased.
At larger energies, instead, above the mobility edge $\varepsilon_{\rm MBL}$, 
$D_1$ and $D_2$ decay to zero in the large $N$ limit. 

These results support the existence of two distinct non-ergodic regions of the phase diagram: a delocalized multifractal phase ($0 < D_1,D_2 < 1$) at intermediate energy density, where eigenstates occupy a volume that diverges, yet is exponentially smaller than the total Hilbert space, and
a Anderson localized phase ($D_1,D_2 \to 0$), where eigenstates are exponentially localized in the Hilbert space and occupy a finite, $N$-indepentent volume on the hypercube.
We have repeated the same analysis for $\Gamma=0.1$, finding similar results. 

Note, however, that the cavity approach does not allow one to determine sharply the phase boundaries between the three phases because the numerical results are only available for systems of moderate size, $N\lesssim 50$, and the asymptotic values of the fractal dimensions cannot be firmly established, especially in the vicinity of the transition line between the bad metal and the MBL phases (see, e.g., the data for $\varepsilon = 0.2$ of Fig.~\ref{fig:D1D2}).
For $\Gamma=0.2$ the MBL mobility edge within the cavity approximation is estimated within the interval $\varepsilon_{\rm MBL} \in (0.15,0.19)$, see fig.~\ref{fig:lev}, which is in good agreement with the estimation of Sec.~\ref{sec:RP}, Eq.~(\ref{eq:rp-f1}) and Fig.~(\ref{fig:RP-micro-phase}).

\subsection{Level statistics}

\begin{figure}
\includegraphics[width=0.49\textwidth]{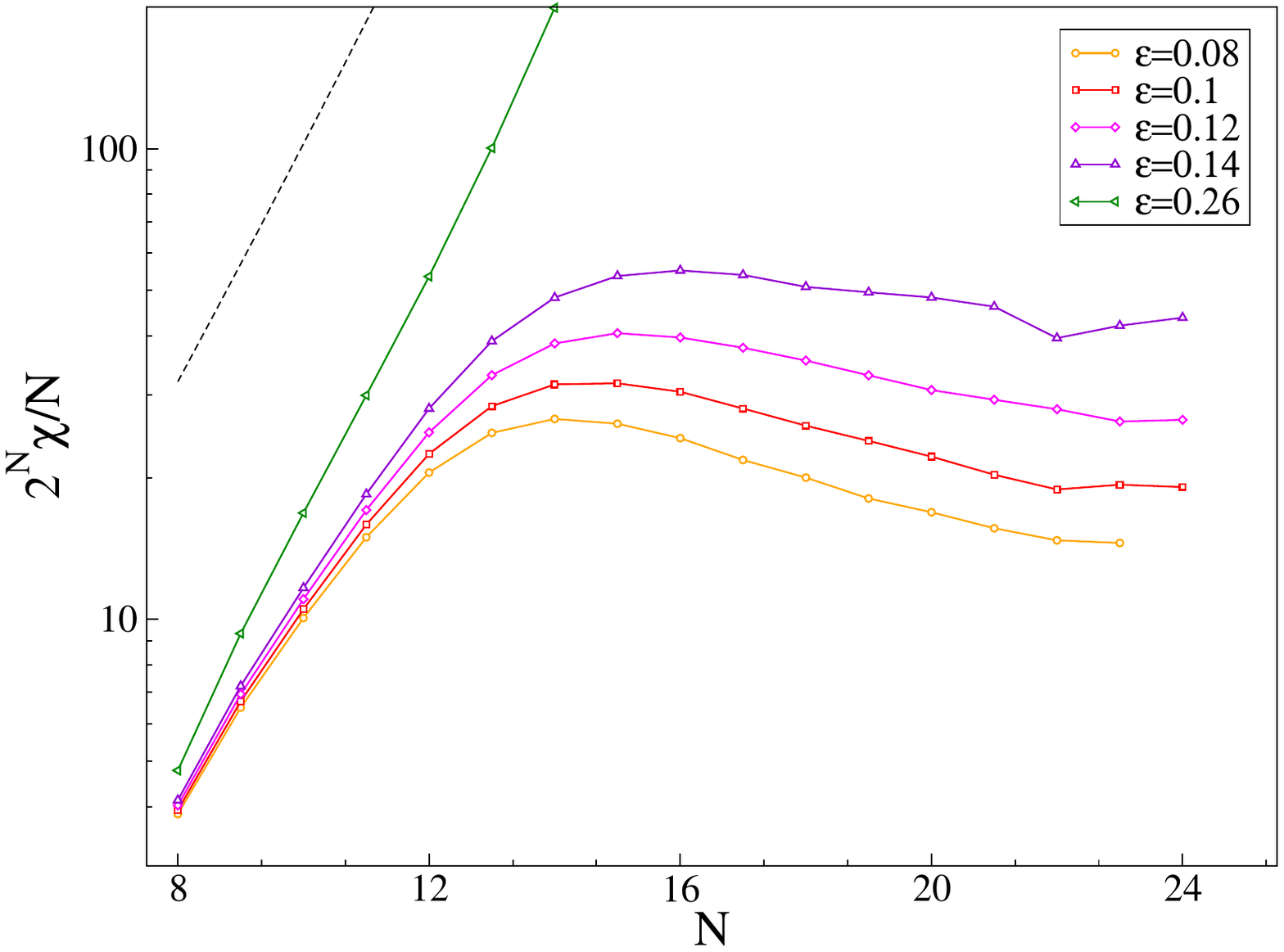}
\vspace{-0.2cm}
\caption{\label{fig:chi} Level compressibility (obtained from the C-BP approach with $n=2$) on the scale of the mean level spacing rescaled by the GOE asymptotic behavior, $2^N \chi(\varepsilon;2 c \delta)/N$, as a function of $N$ for $\Gamma=0.2$ and several values of $\varepsilon$. The black dashed line corresponds to $\chi=1$ (Poisson statistics).}
\end{figure}

A natural question that arises concerns the statistics of the energy levels in the multifractal phase.
In fact, in analogy with the RP model it is reasonable to expect that in the mixed phase the level statistics should be described by the Gaussian orthogonal ensemble (GOE) ensemble on the scale of the mean level spacing, while it might crossover to a different, possibly non-universal behavior, 
on a larger energy scale ($\propto \tilde{\cal V}^{D_2-1}$) which goes to zero exponentially with $N$ but stays much larger than $\delta$.\cite{Kravtsov2015}
This scenario is also supported by general arguments based on the convergence of the Dyson Brownian Motion to its stationary GOE distribution.\cite{Erdos2013DBM,Facoetti2016}
In order to check this idea we have analyzed the behavior of the level compressibility $\chi (\varepsilon;\omega)$ for the number of energy
levels inside the interval $[N\varepsilon-\omega/2, N\varepsilon + \omega/2]$,\cite{Mehta-book} which display different scaling behaviors for the ergodic, localized, and multifractal states.\cite{BiroliTarzia-Review,Metz2017,Altshuler1988,Chalker1996,Bogomolny2011,Mirlin2000Review}
 The number of energy levels inside an energy interval of width $\omega$ (and centered around $N \varepsilon$) is defined as 
	${\cal L} (\varepsilon;\omega) = \int_{\varepsilon-\omega/2}^{\varepsilon + \omega/2} \sum_{\alpha=1}^{2^N} \delta(E^\prime - E_\alpha) \, \textrm{d} E^\prime$.
The level compressibility is then the ratio between the variance of ${\cal L} (\varepsilon; \omega)$ and its average:\cite{Mehta-book}
\[
	\chi(\varepsilon; \omega) = \frac{\overline{({\cal L} (\varepsilon; \omega))^2} - \overline{{\cal L} (\varepsilon; \omega)}^2}{\overline{{\cal L} (\varepsilon; \omega)}} \, ,
\]
where $\overline{\cdots}$ denotes the average over the disorder.
In the diffusive regime of the standard ergodic metallic phase, described by the Wigner-Dyson statistics, energy levels strongly repel each other, and the
variance scales as $\overline{({\cal L} (\varepsilon; \omega))^2} - \overline{{\cal L} (\varepsilon; \omega)}^2 \propto \ln \overline{{\cal L} (\varepsilon; \omega)}$.\cite{Mehta-book}
The level compressibility thus vanishes as $\chi(\varepsilon, \omega) \propto N \ln 2/2^N$ for large $N$.
Conversely, in the localized phase energy levels are thrown as random points on a line and are described by a Poisson distribution.
Hence $\overline{({\cal L} (\varepsilon; \omega))^2} - \overline{{\cal L} (\varepsilon; \omega)}^2 = \overline{{\cal L} (\varepsilon; \omega)}$ and $\chi(\varepsilon; E) \to 1$ for $N \to \infty$.
Finally, for non-ergodic multifractal states the variance of the number of energy levels inside an interval should scale linearly with the 
average,\cite{Altshuler1988,Chalker1996,Bogomolny2011,Mirlin2000Review} at least in simplest scenarios, and $\chi(\varepsilon; \omega)$ is expected to converge to a (system-dependent) constant between $0$ and $1$ in the 
large $N$ limit.
In the following for simplicity we will only focus on the behavior of the level compressibility when the energy interval $\omega$ is taken of the order of the mean level spacings. In particular we will set $\omega = 2 \eta = 2 c \delta$, where $\eta$ is given in Eq.~(\ref{eq:eta}) and $c = 64$.\footnote{We have checked that varying $c$ from $16$ to $128$ do not modify the results.}

As shown in Refs.~[\onlinecite{BiroliTarzia-Review}] and~[\onlinecite{Metz2017}] a simple spectral representation of ${\cal L} (\varepsilon;\omega)$ 
can be achieved in the framework of the C-BP approach, 
in terms of the resolvent matrices defined on the clusters of the hypercube and of the cavity resolvent matrices defined on the edges between the clusters:
\begin{equation} \label{eq:NE2}
\begin{aligned}
{\cal L} (\varepsilon; \omega) =& \frac{1}{\pi} \! \lim_{\eta \to 0^+} \! \bigg \{ \sum_{p=1}^{2^{N-n}} \big[ \Psi_p ( z_+ ) - \Psi_p (  z_- ) \big]\\
& \qquad \,\,\,\,\,\, + \sum_{\langle p, q \rangle} \big[ \varphi_{p \leftrightarrow q} ( z_+ ) - \varphi_{p \leftrightarrow q} (  z_- ) \big] \bigg \} \, ,
\end{aligned}
\end{equation}
where $z_{\pm} = N \varepsilon \pm \omega/2 + i \eta$, 
the angle $\Psi_p (z)$ is defined as the phase of 
${\rm det} \, {\cal G}_p (z)$, ${\rm det} \, {\cal G}_p (z) = | {\rm det} \, {\cal G}_p (z) | e^{i \Psi_p (z)}$, and the angle $\varphi_{p \leftrightarrow q} ( z)$ is defined as the phase of 
${\rm det} \, ({\cal I}_s - \Gamma^2 G_{q \to p} (z)  G_{p \to q} (z))$
(we have chosen here to put the branch-cut in the complex plane along the negative real axis).

In order to analyze the scaling properties of the level compressibility
we then just need to compute the average of ${\cal L} (\varepsilon; \omega)$ and its fluctuations over many independent
realizations of the random energies of the hypercube. The scaling behavior of $\chi$ when $\omega$ is taken on the scale of the mean level spacing, $\omega = 2 c \delta$, are shown in fig.~\ref{fig:chi}, where we plot the compressibility (divided by the GOE asymptotic) 
versus $N$ for $\Gamma=0.2$ and several values energy density in the region of multifractal eigenstates. We observe that $2^N \chi/N$ has a non-monotonic behavior roughly on the same scale as $D_1$ and $D_2$, and seems to approach a finite value at large $N$ which grows as $\varepsilon$ is increased towards the localized phase. For $\varepsilon \gtrsim 0.26$, in the MBL phase, $\chi$ tends instead to $1$ at large $N$, as expected for Poisson statistics. This implies that in the whole multifractal region, on energy scales proportional to $\delta$ the level compressibility goes to zero in the thermodynamic limit as in the diffusive regime of the standard metallic phase.

\section{Discussion} \label{sec:PD}
Above we have presented two complementary approximate strategies to determine the out of equilibrium phase diagram of the QREM. The first approach is based on the FSA and on the mapping to the RP model,\cite{Faoro2019} while the second approach is a generalization of the self-consistent theory of Anderson localization\cite{ATA,Logan2019,RoyLogan2020} adapted to take into account (at least partially) the local structure of the Hilbert space of the QREM.  As discussed more in detail in App.~\ref{app:cavity} the latter possibly provides a quite accurate approximation of local observables, such as the distribution of the LDoS, while the former is expected to yield a better estimation of correlations, since it is able to capture the fact that there is a factorial number of paths connecting two points at large distance in the configuration space.

\begin{figure}
	\includegraphics[width=0.47\textwidth]{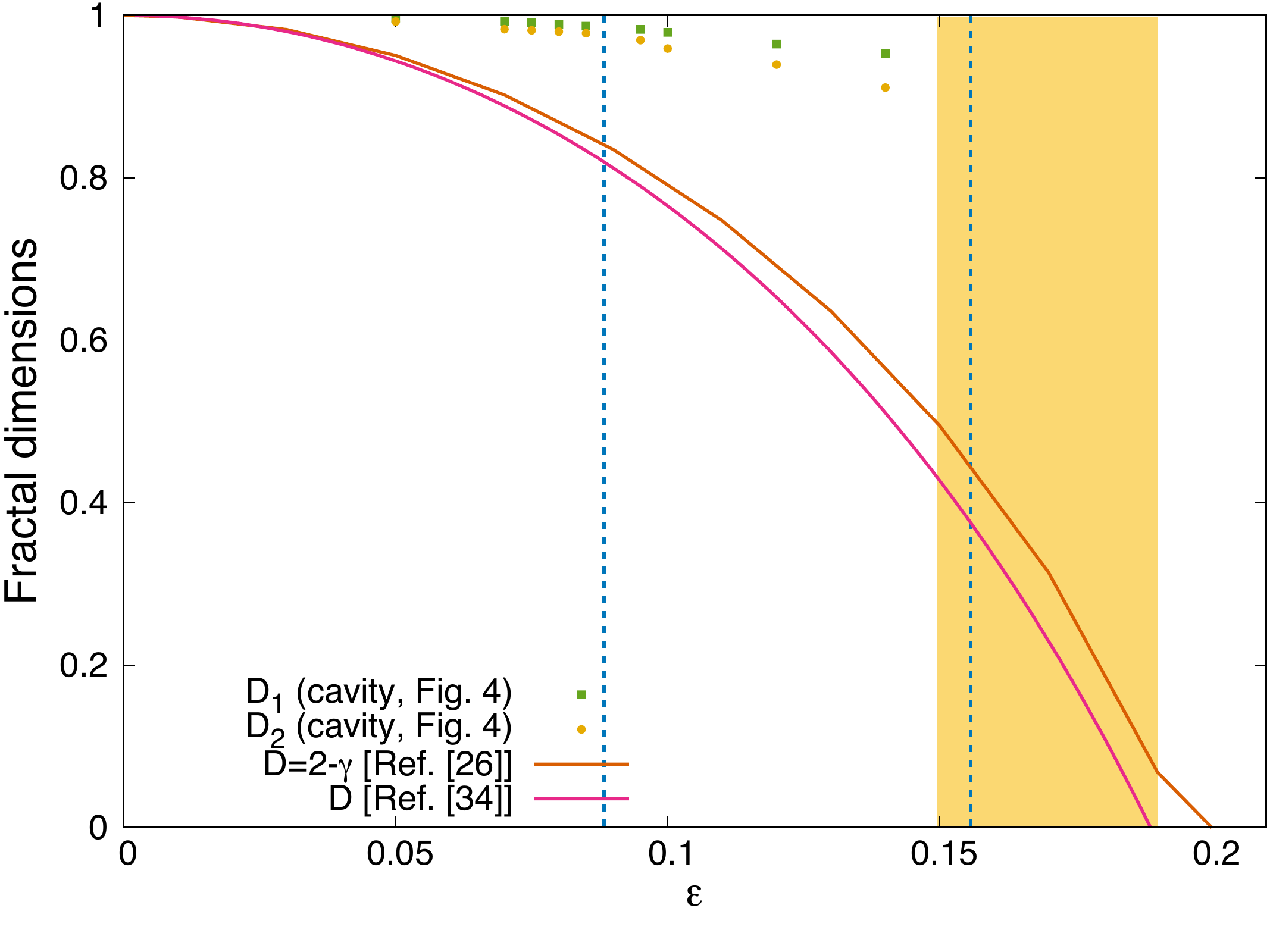}
	\vspace{-0.4cm}
	\caption{\label{fig:lev} Fractal dimensions $D_1$ (green squares) and $D_2$ (yellow circles) obtained from the cavity approximation as a function of $\varepsilon$ for $\Gamma=0.2$, determined by estimating the height of the plateaus of fig.~\ref{fig:D1D2} at large $N$. 
	The yellow shaded region indicates the energy interval within which the numerical results of fig.~\ref{fig:D1D2} suggest that the MBL transition occurs, $0.15 \lesssim \varepsilon_{\rm MBL} \lesssim 0.19$.
	The vertical dashed blue lines show the position of $\eerg \approx 0.0885$ and of $\varepsilon_{\rm MBL} \approx 0.1585$ found in Sec.~\ref{sec:RP} using the FSA and the mapping onto the RP model, Eqs.~(\ref{eq:qp-dyn-f1}) and~(\ref{eq:rp-f2}).
	We also plot the results for the fractal dimension $D$ obtained in Refs.~[\onlinecite{Faoro2019}] (orange line) and [\onlinecite{Smelyanskiy2019}] (magenta line), which predict that $D \to 1$ for $\varepsilon \to 0$ (i.e., $\eerg \to 0$).}
\end{figure}

In this section we discuss and compare more in details the two approaches for what concerns the behavior of the  fractal dimensions $D_1$ and $D_2$ as a function of the energy density $\varepsilon$, that we plot in fig.~\ref{fig:lev}  for $\Gamma=0.2$ in the large-$N$ limit.

Within the analogy between the QREM and the RP model discussed in Sec.~\ref{sec:RP}, the fractal dimension is expected to be equal to one in the ergodic phase, $|\varepsilon| < \eerg$, and to zero in the MBL phase, $|\varepsilon| > \varepsilon_{\rm MBL}$, and is conjectured to decrease from $1$ to $0$ in the intermediate NEE regime (the multifractal spectrum should be obtained as an ``average'' of the effective fractal dimensions over all $x$-sectors).  

We also plot the estimations of Refs.~[\onlinecite{Faoro2019}] (orange line) 
and~[\onlinecite{Smelyanskiy2019}]
where the effective spectral dimension $D$ 
is obtained using similar 
(although probably more accurate) methods to evaluate the amplitude of the tunneling rates $\langle \{ \sigma_a^z \} \vert \Gamma \sigma_b^x \vert  \{ \sigma_a^z \}^\prime \rangle$ between two distant many-body configurations.  The symbols are the cavity-cluster predictions, extracted from the largest size available  when reasonably converged to a plateau (see fig.~\ref{fig:D1D2}). The shaded area indicates the energy interval within which the numerical results of fig.~\ref{fig:D1D2} suggest that the MBL transition should take place.
Due to the limited range of system sizes accessible via the cavity approach, 
we are not able to conclude whether the fractal dimension would continuously go to zero at $\varepsilon_{\rm MBL}$ or rather exhibit a finite jump at the transition.

All approaches agree in indicating the existence of three different phases of the QREM: a fully ergodic regime at low energy density, $|\varepsilon| < \eerg$, a NEE (or bad metal, or dynamical glassy) one at intermediate energy density, $\eerg < |\varepsilon| < \varepsilon_{\rm MBL}$, where the time to reach thermal equilibrium is exponentially large in the system size and eigenvectors are extended but multifractal, corresponding to $0 < D_{1,2} < 1$,  and a fully localized one, with Anderson localized eigenstates, at high energy, $|\varepsilon |> \varepsilon_{\rm MBL}$ where $D_{1,2} \rightarrow 0$.

However, some quantitative differences also emerge between these three approximations. According to the FSA calculation of Sec.~\ref{sec:RP}, the ergodic region extends up to a finite energy (as also recently suggested by the numerical results of Ref.~[\onlinecite{Parolini2020}]), while the approach of Refs.~[\onlinecite{Faoro2019}] and [\onlinecite{Smelyanskiy2019}] predicts instead that $\eerg\to 0$. The cavity approximation indicates that if  $\eerg$ is finite, it is significantly smaller than the estimation of Eqs.~(\ref{eq:rp-ergo}) and~(\ref{eq:rp-f2}). As we are going to see in the next section, an argument based on a simplified solution of the cavity equations, equivalent to an auxiliary Anderson model in unconventional thermodynamic limit, also seem to suggest that $\eerg$ might indeed squeeze to zero energy in the thermodynamic limit (as $\sqrt{\ln N / N}$ in the large $N$ limit).\cite{Smelyanskiy2019}


The other difference is that within the cavity approach the fractal dimensions $D_1$ and $D_2$ might possibly exhibit an abrupt jump from a finite value smaller then one to zero at the transition between the nonergodic delocalized and fully localized regime, while the mapping to the RP model indicates that if one identifies $D = 2 - \gamma$,\cite{Kravtsov2015} the fractal dimensions should vanish continuously at the MBL mobility edge, as also found in Refs.~[\onlinecite{Faoro2019}] and [\onlinecite{Smelyanskiy2019}].
These are still open questions for future investigations.

\section{Auxiliary Anderson Models and Unconventional Thermodynamic limit} \label{sec:toy}

In this section we further simplify the quantum cavity analysis and introduce a family of auxiliary ``toy'' Anderson tight-binding models on a Bethe lattice with connectivity $N\gg 1$ 
where the volume $\vol$ of the system is treated as an independent parameter from $N$ and taken equal to infinity from the start (see also Refs.~[\onlinecite{Logan2019,RoyLogan2020,RoyLogan2020-2}] and~[\onlinecite{RoyLazarides2020}] for similar approaches in the context of MBL).
The basic idea behind this procedure is that in the original Anderson model on the Hypercube (see section~\ref{sec:HS}) the scaling of the number of sites and number of neighbors with respect to the size $N$ of the original QREM is remarkably different, the former being exponential $\vol = 2^N$ while the latter is linear $k = N$. As such one could hope to get some insight on the solution of the quantum cavity equations by taking the volume to infinity first, possibly providing an estimation for the transition line between the fully ergodic phase and the multifractal bad metal one. 

Concretely, we consider a hybrid version of the model~(\ref{eq:H}) where the total number of sites of the lattice is sent to infinity from the start keeping $N$ fixed. Hence, for any given choice of $\Gamma$ and $\varepsilon$, this leads to a family of tight-binding Anderson model parametrized by the connectivity $N$, with random on-site disorder of standard deviation $\sqrt{N/2}$, given by Eq.~(\ref{eq:PE}). The advantage of this procedure is that now for any choice of $\Gamma$, $\varepsilon$, and $N$, the imaginary regulator can be taken as infinitesimally small (since the mean level spacing vanishes in the thermodynamic limit) and we can study whether the system is in the localized or in the extended phase with standard techniques.

We start by determining the mobility edge $\varepsilon_{\rm loc} (N) = E_{\rm loc}(N)/N$ of the auxiliary models by computing the Lyapunov exponent which describes the evolution of the imaginary part $\Delta$ 
of the self-energy $\Sigma$, once the iteration relations~(\ref{eq:recursion}) have been linearized.\cite{ATA} At a given order $n$ of the cluster expansion, the  (cavity) self-energy on a cluster $p$ of $s=2^n$ sites (in absence of the $2^n$ edges with one of the neighboring clusters $q$) is a $s \times s$ matrix defined as:
\[
\Sigma_{p \to q} = S_{p \to q} + i \Delta_{p \to q} = {\cal H}_p - z {\cal I}_s - G^{-1}_{p \to q} \, ,
\]
where ${\cal H}_p$ is the Hamiltonian~(\ref{eq:H}) acting on the sites of the cluster, and ${\cal I}_s$ is the $s \times s$ identity matrix.
In the localized phase its imaginary part 
vanishes for $\eta \to 0^+$.
Hence, in the thermodynamic limit and close to the localization transition, one can take the limit $\eta \to 0^+$ from the start and linearize the recursive equations~(\ref{eq:recursion}) with respect to $\Delta$: 
\begin{eqnarray}
 \label{eq:SigmaLinR}
\left[ S_{p \to q_k} \right]_{uv} & = &
\Gamma^2 \!\!\! \sum_{q_l \in \partial p / q_k} \left[ {\cal H}_{q_l} - N \varepsilon \, {\cal I}_s - S_{q_l \to p} \right]^{-1}_{uv} \, , \\
\label{eq:SigmaLin}
\left[\Delta_{p \to q_k}\right]_{uv} & = &
\Gamma^2 \!\!\! \sum_{q_l \in \partial p / q_k} \sum_{w,y=1}^s \left[ {\cal H}_{q_l} - N \varepsilon \, {\cal I}_s - S_{q_l \to p} \right]_{uw}^{-1} \\
\nonumber 
& & \qquad \qquad \times \left[ \Delta_{q_l \to p} \right]_{wy} \left[ {\cal H}_{q_l} - N \varepsilon \, {\cal I}_s - S_{q_l \to p} \right]_{yv}^{-1} \, .
\end{eqnarray}
Since $S_{p \to q}$ and $\Delta_{p \to q}$ are random matrices, 
these equations naturally lead to functional self-consistent equations on their probability distribution (see also App.~\ref{app:analytic}), which can be solved with arbitrary numerical precision using a population dynamics algorithm\cite{BiroliSemerjianTarzia,BiroliTarzia-Review} for each value of $\Gamma$, $\varepsilon$ and $N$.

The Lyapunov exponent $\Lambda$ describes the exponential growth or the exponential decay of the imaginary part of the diagonal elements of the self-energy with the number of recursion steps $\recK$ as: $\Delta_{\rm typ} \propto e^{\Lambda \recK}$. 
However, in the delocalized phase after few recursions $\Delta_{\rm typ}$  becomes of order $1$ and the exponential behaviour is lost.
To circumvent this problem we follow Ref.~[\onlinecite{Altshuler2016}] and add an additional ``inflationary'' step to the recursion:
\begin{enumerate}
	\item We reach to the stationary distribution of the real part of the self-energies, Eq.~(\ref{eq:SigmaLinR});
	\item  We initialize the imaginary parts of the self-energy to  very small values, e.g. $\Delta_{\rm typ} = \theta = 10^{-24}$; 
	\item We execute an iteration step using Eq.~(\ref{eq:SigmaLin}) and update the whole population;
	\item We compute $\Delta_{\rm typ}$ after the iteration; 
	\item We multiply all the imaginary parts by $e^{\hat{\Lambda}} = \theta / \Delta_{\rm typ}$, such that  the typical value of $\Delta$ remains equal to $\theta$ at each iteration step;
	\item We go back to step 4.
\end{enumerate}
An estimate of the Lypaunov exponent $\Lambda$ can be then obtained as  the average value of $\hat{\Lambda}$ at stationarity (in principle one should take  $\theta \to 0$ and consider the  infinite population size limit).

\begin{figure}
\includegraphics[width=0.46\textwidth]{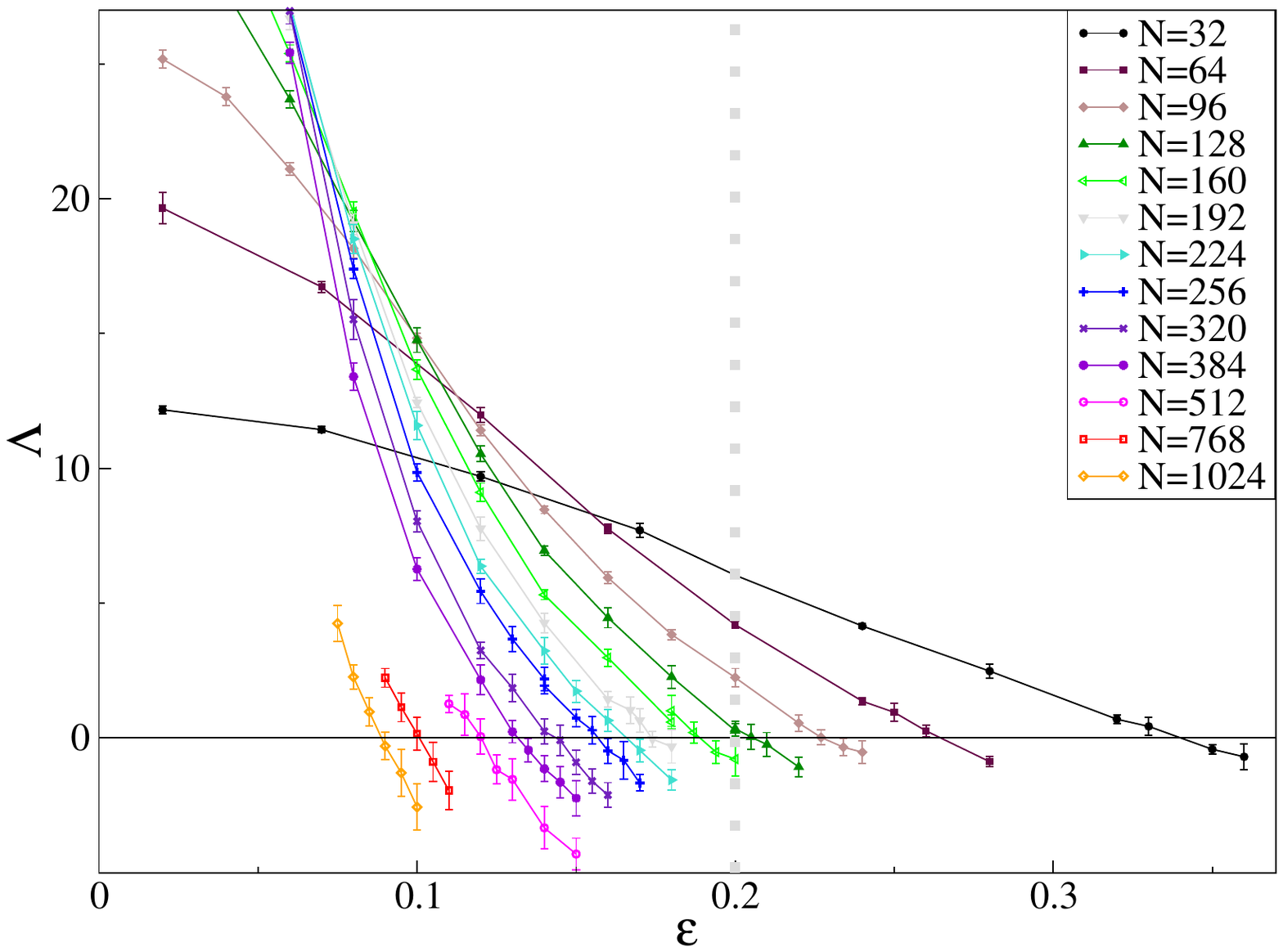}
\vspace{-0.2cm}
	\caption{\label{fig:lyap} Lyapunov exponent $\Lambda$ as a function of the energy $\varepsilon = E/N$ of the hybrid Anderson tight-binding models when ${\cal V}$ is sent to infinity at fixed $N$, for several values of $N$ ranging from $32$ to $1024$ and for $\Gamma=0.2$. The results are obtained within the cluster expansion with $n=2$. The vertical gray dotted line corresponds to the prediction of the FSA for the localization threshold $\varepsilon_{\rm loc} = \Gamma$ (see App.~\ref{app:analytic}).}
\end{figure}

In fig.~\ref{fig:lyap} we report the results of an accurate numerical computation of the Lyapunov exponent for $\Gamma=0.2$ and for several values of the the energy density $\varepsilon$ and of the connectivity $N$, performed with populations' size $M$ ranging from $2^{25}$ to $2^{27}$, and with $\theta$ from $10^{-16}$ to $10^{-24}$ (and for $n=2$).

\begin{figure}
\includegraphics[width=0.46\textwidth]{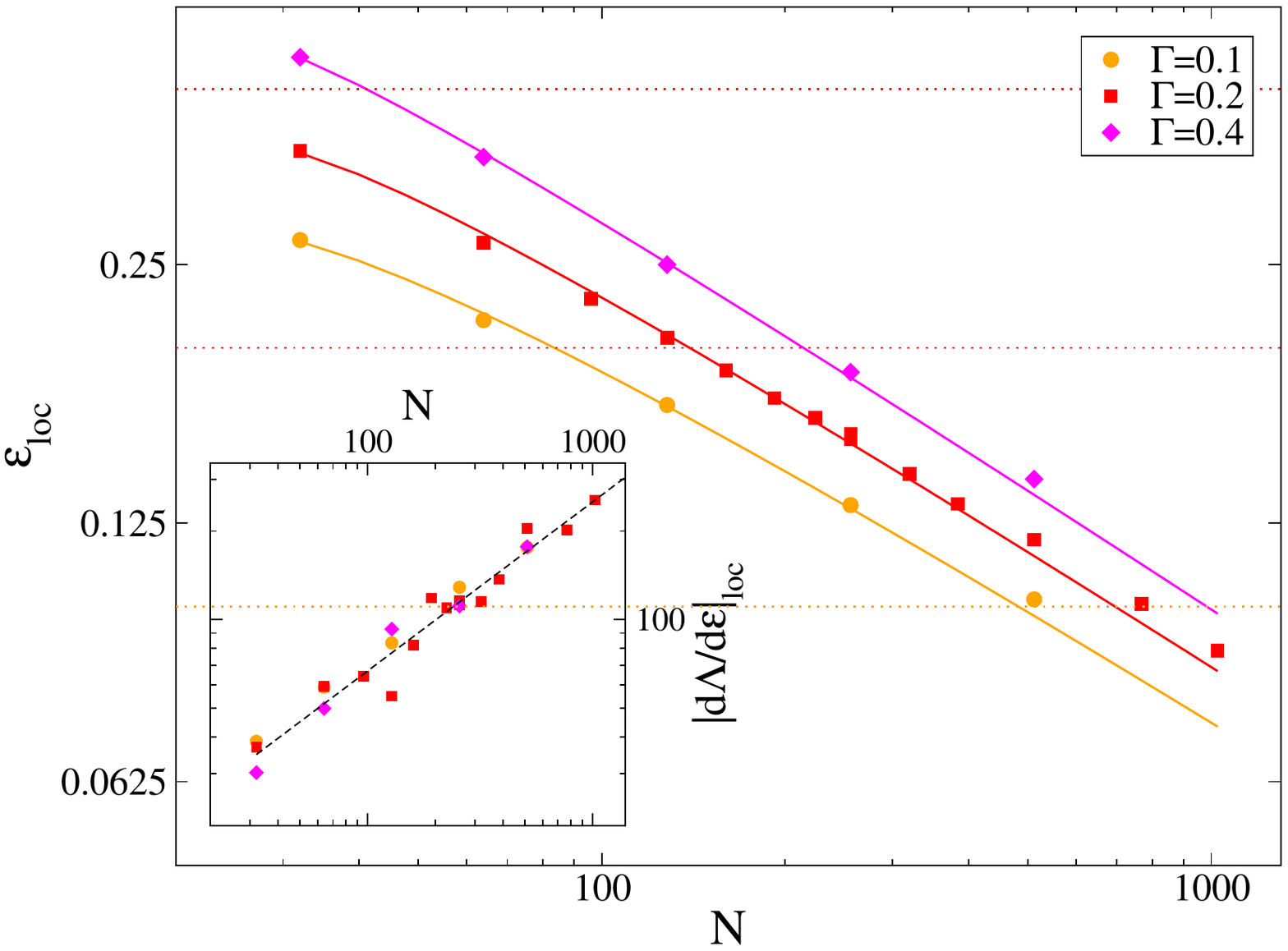}
\vspace{-0.2cm}
	\caption{\label{fig:zero} Main panel: Mobility edge $\varepsilon_{\rm loc} (N)$ (such that $\Lambda(\varepsilon_{\rm loc})=0$) of the family of models~(\ref{eq:H}) when the thermodynamic limit is taken from the start, as a function of $N$ for three different values of $\Gamma$ (and for $n=2$). The continuous curves correspond to the analytical prediction of Eq.~(\ref{eq:mobility}), with the fitting parameters $c_1$ and $c_2$ smoothly varying with $\Gamma$ as $c_1\approx -11.51$, $-9.27$, and $-6.62$, and $c_2\approx 3.25$, $4.54$, and $6.5$ for $\Gamma=0.1$, $0.2$, and $0.4$ respectively. The horizontal dotted lines correspond to the prediction of the FSA, $\varepsilon_{\rm loc} = \Gamma$. Inset: Slope of the Lyapunov exponent computed at $\varepsilon_{\rm loc}$ as a function of $N$ for the same values of $\Gamma$ as in the main panel. The black dashed line is a pwer-law fit of the data as $|{\rm d} \Lambda / {\rm d} \varepsilon\vert_{\varepsilon_{\rm loc}} \simeq A \, N^\gamma$, with $\gamma \approx 0.57$.}
\end{figure}

One observes that the critical energy ``density'' $\varepsilon_{\rm loc} (N)$ at which the Lyapunov exponent vanishes 
slowly but continuously decreases as $N$ is increased.
In Fig.~\ref{fig:zero} we plot the dependence of $\varepsilon_{\rm loc} (N)$ as a function of $N$ for three different values of $\Gamma$: We find a similar behavior for all values of the transverse field, at least in the region of the phase diagram where the physical properties of the QREM are dominated by the random term, $\Gamma < \sqrt{2}/2$.\cite{Goldschmidt1990,Jorg2008} In particular, we observe that $\varepsilon_{\rm loc} (N)$ seems to vanish in the large $N$ limit as $\sqrt{\ln N/N}$. Consistently, we find that the slope $|{\rm d} \Lambda / {\rm d} \varepsilon\vert_{\varepsilon_{\rm loc}}$ around $\varepsilon_{\rm loc}$ grows roughly as $\sqrt{N}$ (inset).

This behavior can be understood in terms of the analytic computation, carried out in App.~\ref{app:analytic} in full details, of the largest eigenvalue of the integral operator associated to the self-consistent equation for $Q(S,\Delta)$ in the large-connectivity limit and for $n=0$ (i.e., in the standard single-site Bethe approximation, when the underlying lattice is taken as a RRG of connectivity $N$), which yield:\cite{Bapst2014,ATA,Parisi2019AndersonBethe}
\begin{equation} \label{eq:mobility}
\varepsilon_{\rm loc} 
= \left [ \frac{\ln \left [ \sqrt{N \Gamma^2 / \pi} \,\left(  \ln \left( N / \Gamma^4 \right)  + c_1 \right) \right] + c_2}{N} \right]^{1/2},
\end{equation}
where $c_1$ is  
a real constant of $O(1)$ which only depend on $\Gamma$, and $c_2$ can be expressed in terms of the solution $\mu_\star$ of the self-consistent equations~(\ref{eq:A-mu}) for the real part of the self-energy.
The predictions of this equations are plotted in Fig.~\ref{fig:zero} on top of the numerical points, showing a good agreement with the numerical results.

It is interesting to compare the asymptotic behavior at large $N$ of the localization threshold $\varepsilon_{\rm loc}$ found here for the family of auxiliary models with the large connectivity limit of the standard Anderson model on the Bethe lattice.\cite{Bapst2014,ATA,Parisi2019AndersonBethe} In fact, for Bethe lattices of connectivity $k+1$ and on-site random energies taken from a box distribution of width $W$, in the large $k$ (and large disorder) limit the localization transition 
(at zero energy) takes place at disorder $W_L$ given by:
\begin{equation} \label{eq:anderson}
4 \, \rho \ln (W_L) = \frac{1}{k} \, ,
\end{equation}
where $\rho$ is the density of state in the middle of the spectrum which, at strong disorder, is just given by $\rho(0) \simeq 1/W$.
In order to translate this relation to our case, assuming that
that $\rho(E) \approx P(E)$, one finds that the exponential dependence on $N$ of the DoS is exactly canceled for $E_{\rm loc}=N \varepsilon_{\rm loc}$ given by Eq.~(\ref{eq:mobility}):
\[
\rho (\varepsilon_{\rm loc}) \approx \frac{e^{-c_2}}{N \Gamma \ln (N / \Gamma^4)} \, .
\]
(We have neglected the constant $c_1$ for simplicity.)
The variance of the random energies of the QREM scales as $\sigma_E^2 = N/2$, which leads us to identify the effective disorder as 
$W \approx \sqrt{6 N}$. Thus, 
from Eq.~(\ref{eq:anderson}) one gets:
\[
\frac{4 \, e^{-c_2} \ln \sqrt{6 N}}{N \Gamma \ln (N / \Gamma^4)} \approx \frac{1}{N} \, ,
\]
which is satisfied for $c_2 = \ln(2/\Gamma)$.

Going back now to the QREM and to its Hilbert space formulation~(\ref{eq:H}) 
defined on {\it finite} boolean hypercubes of ${\cal V} = 2^N$ sites, we argue that the mobility edges of the auxiliary models provide asymptotically in the large $N$ limit an estimation for the transition line between the fully ergodic phase and the delocalized but multifractal one.
The argument goes as follows: 
Consider a wave-function defined on the $N$-dimensional  hypercube which decays exponentially over a finite, $N$-independent length $\xi$. This corresponds to a multifractal many-body state which occupies roughly $\xi^N$ sites of the Hilbert space, with a fractal dimension $D \sim \log_2 \xi$.
In the original many-body setting, when the number of spins $N$ is sent to infinity, the volume of the Hilbert space ($2^N$) and the connectivity of the hypercube ($N$) diverge concomitantly, and such wave-function will preserve its multifractal nature at all $N$. If, instead, the volume of the Hilbert space is sent to infinity while the connectivity is kept fixed, as in the auxiliary model, then such wave-function will appear as genuinely Anderson localized as it occupies a {\it finite} volume. In this sense, it is natural to argue that the apparent $N$-dependent Anderson localization transition of the family of auxiliary models may in fact captures the transition from ergodic to non-ergodic eigenstates of the original problem.

On the other hand, the FSA analysis of the linearized recursion relations of the auxiliary models on the Bethe lattice, which consists in neglecting the real part of the
self-energy in the denominators of Eqs.~\eqref{eq:SigmaLinR} and~\eqref{eq:SigmaLin}, simply predicts (again in the simplest $n=0$ setting) that 
$\varepsilon_{\rm loc}^{\rm FSA} = \Gamma$, 
irrespectively of $N$ (see App.~\ref{app:analytic} for a detailed calculation).
This is the same result for the many-body mobility edge of the QREM at the lowest order in $\Gamma$~\cite{Laumann2014}. We argue that the localization threshold predicted by the FSA does not depend on whether the ${\cal V} \to \infty$ limit is taken before the $N \to \infty$ one or not. In fact, the FSA only keeps the leading-order contribution to the wave-function amplitude at each site, and determine the convergence of the perturbative expansion by counting the relative number of resonances found at a given distance, compared to the total number of sites accessible at such distance.
In this sense, this approximation captures the transition from the Anderson-localized regime (where the perturbative expansion is convergent, the eigenstates are weakly-dressed single configurations of spins and occupy a finite volume on the hypercube) to a delocalized regime (where the perturbative expansion does not converge, and resonances can be found at arbitrary large distances) irrespectively of the multifractal nature of the eigenstates.
These arguments thus suggest that, while $\varepsilon_{\rm loc}^{\rm FSA}$ gives a rough estimate of the mobility edge between the MBL phase and the NEE phase,\cite{Laumann2014} $\varepsilon_{\rm loc}$ given in Eq.~(\ref{eq:mobility}) provides an estimation of the transition between the fully ergodic phase and the delocalized non-ergodic one:
\begin{equation} \label{eq:transitions}
\begin{aligned}
\varepsilon_{\rm MBL} & \approx \Gamma \, , \\
\eerg & \approx  \left [ \frac{\ln \sqrt{N \Gamma^2 / \pi} + O (\ln (\ln  N))}{N} \right]^{1/2} \, .
\end{aligned}
\end{equation}
According to this interpretation,  
fully ergodic eigenstates of the QREM are only found in a narrow energy window around $|\varepsilon| < \eerg$, which concentrate around zero in the thermodynamic limit (in agreement with Refs.~[\onlinecite{Faoro2019}] and~[\onlinecite{Smelyanskiy2019}].
Yet, the fraction of ergodic eigenfunctions at large $N$ is approximately given by $\int_{-N \eerg}^{N \eerg} \rho(E) \, {\rm d} E 
\approx 1 - C/\sqrt{N(\ln N)^3}$, with $\rho(E) \approx e^{-E^2/N} / \sqrt{\pi N}$, and $C$ being a constant of order $1$ which depends on $\Gamma$. Hence, although $\eerg \to 0$, due to the scaling of the many-body energies with $N$, only a fraction of order $1/\sqrt{N}$ of the $2^N$ eigenstates are delocalized but non-ergodic. Furthermore, a finite value of the mobility edge, $\varepsilon_{\rm MBL} \approx \Gamma$, only corresponds to an exponentially small fraction of Anderson-localized eigenstates in the tails of the DoS.

\section{Conclusions and perspectives} \label{sec:conclusions}

In this work we have revisited the dynamical phase diagram of the QREM, using a complementary set of approaches, the FSA coupled to a mapping to the RP model\cite{Faoro2019} and the self-consistent theory of localization~\cite{ATA}  extended to include the local Hilbert space structure of the QREM.\cite{Logan2019,RoyLogan2020}
While the FSA is expected to yield a better estimation of correlations and large distance physics  the latter provides a quite accurate approximation of local observables. These approaches provide a qualitatively similar scenario for the phase diagram in the energy-transverse field plane, namely the existence of three dynamical phases: a fully ergodic delocalized one, an intermediate non-ergodic extended regime with multifractal behavior, and an Anderson localized one. For what concerns the quantitative features of the phase diagram and the properties of the three phases there remain however many open questions. The FSA and our RP mapping seem to suggest that ergodic delocalized states exist in an entire region around zero energy density (see also \cite{Baldwin2018}), while the analysis of the quantum cavity equations suggest that if such a region exists it is much narrower in energy, in agreement with Refs.~[\onlinecite{Faoro2019}] and~[\onlinecite{Smelyanskiy2019}]. 
On the other hand, an approximate analytic solution of the cavity equations, corresponding to an auxiliary Anderson model on a Bethe lattice where the connectivity is sent to infinity after the thermodynamic limit is taken, mimicking the exponential scaling of the number of sites of the hypercube, points toward a threshold energy for full delocalization squeezing to zero in the thermodynamic limit.\cite{Faoro2019,Smelyanskiy2019} 

It is worth stressing upon concluding that certain features of the QREM make it very peculiar, and produce some specific and unique features compared to other generic interacting disordered models such as $1d$ disordered spin chains or mean-field spin glass models. First among all the absence of correlations between the many-body energies $E_i$'s and the spin configurations $\{ \sigma_a^z \}$. The other unique feature of the QREM is the fact that in the frozen glassy phase,  $T<T_K$, the Edwards-Anderson order parameter is equal to one, implying that essentially no spin can be flipped with respect to the initial state. This property implies that in the MBL phase of the QREM many-body wavefunctions are genuinely Anderson localized and occupy a {\it finite} volume in the Hilbert space, while for generic interacting models one expects that the volume occupied by many-body eigenstates in the configuration space is subexponentially large due to the presence of a finite fraction of active spins.\cite{Mace2018,Tarzia2020} This makes interesting and worth pursuing the investigation using techniques developed here of other disordered mean field models.
\begin{acknowledgments}
We warmly thank L. Ioffe for helpful discussions.
This work was partially supported by the grant from the Simons Foundation (\#454935 Giulio Biroli). 
DF was partially supported by the EPSRC (CANES, EP/L015854/1) and the European Research Council (ERC) under the European Union's Horizon 2020 research and innovation programme (grant agreement n$^{\circ}$~723955 - GlassUniversality).
\end{acknowledgments}

\appendix
\newpage

\begin{figure}
	\vspace{-0.2cm}
	\includegraphics[width=0.48\textwidth]{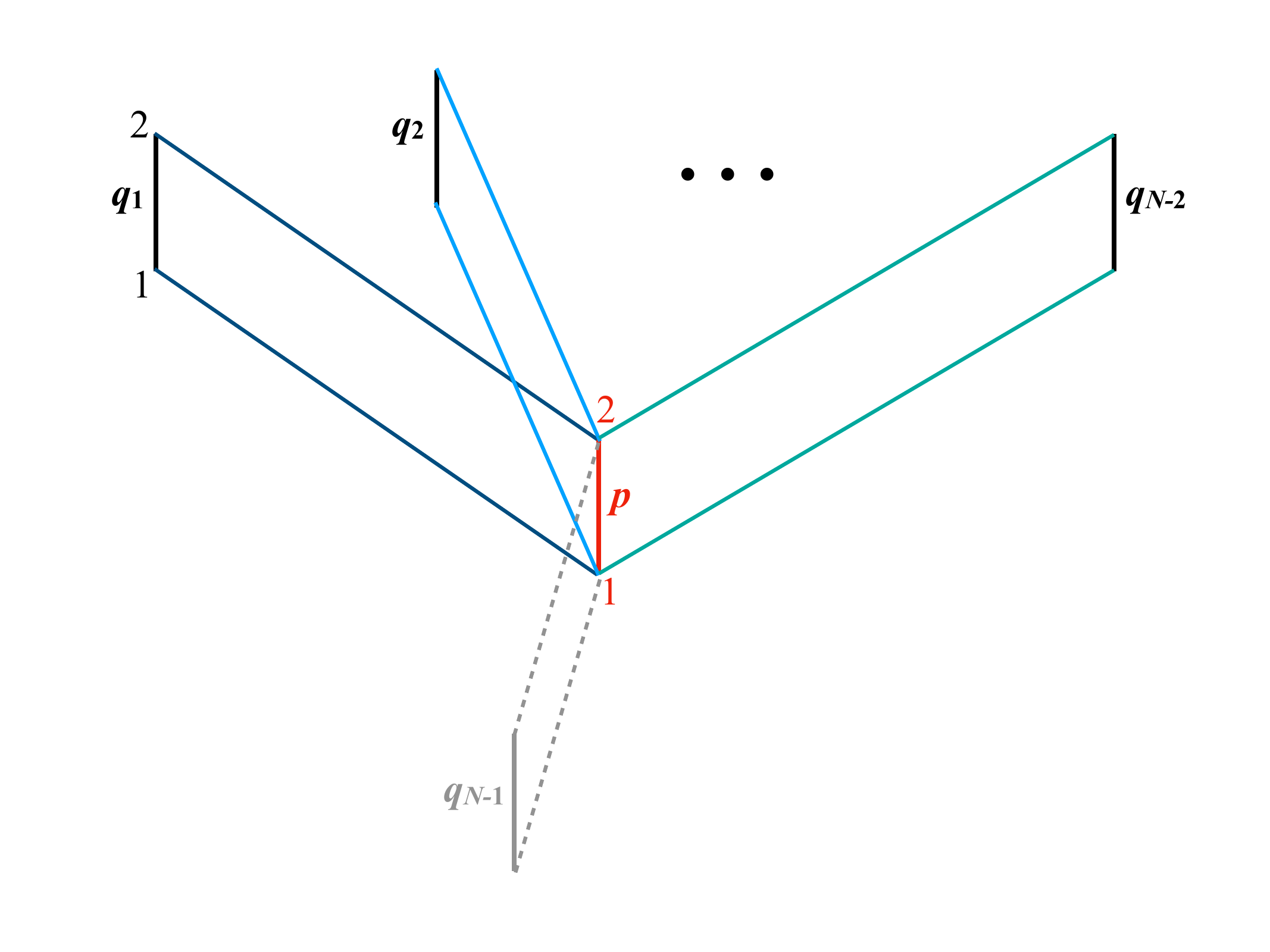}
	\caption {Schematic representation of the recursion step which yields the self-consistent equations for the $2 \times 2$ elements of the (cavity) resolvent matrix for clusters of $2$  sites 
		($n=1$).}
	\label{fig:edges}
\end{figure}
\section{Recursion relations for the matrix elements of the resolvent within the cluster approximation} \label{app:clusters}

For $n=1$ the clusters are simply made by two sites (corresponding to two spin configurations which differ by a single spin flip) connected by an edge (Fig.~\ref{fig:edges}). The cavity resolvent matrix on such cluster can then just be parametrized by three complex numbers:
\[
G_{p \to q} = \left (
\begin{array}{cc}
g_{1}^{p \to q} & g_{12}^{p \to q} \\
g_{12}^{p \to q} & g_{2}^{p \to q}
\end{array}
\right) \, .
\]
The matrix elements of the Hamiltonian~(\ref{eq:H}) on the sites of the cluster are:
\[
{\cal H}_c = \left (
\begin{array}{cc}
-E_1 & -\Gamma \\
-\Gamma & -E_2
\end{array}
\right) \, .
\]
Thus, Eq.~(\ref{eq:recursion}) becomes:
\[
\begin{aligned}
\frac{g_{2}^{p \to q_k}}{{\rm det} \, G_{p \to q_k}}
& = -E_1 - z - \Gamma^2 \!\!\!\! \sum_{q_l \in \partial p / q_k}  \!\!\!\! g^{q_l \to p}_{1}  \, , \\
\frac{g_{1}^{p \to q_k}}{{\rm det} \, G_{p \to q_k}}
& = -E_2 - z - \Gamma^2 \!\!\!\! \sum_{q_l \in \partial p / q_k}  \!\!\!\! g^{q_l \to p}_{2}  \, , \\
\frac{g_{12}^{p \to q_k}}{{\rm det} \, G_{p \to q_k}}
& = -\Gamma + \Gamma^2 \!\!\!\! \sum_{q_l \in \partial p / q_k}  \!\!\!\! g^{q_l \to p}_{12}  \, , 
\end{aligned}
\]
where ${\rm det} \, G_{p \to q_k} = g_{1}^{p \to q_k} g_{2}^{p \to q_k} - (g_{12}^{p \to q_k})^2$.
From Eq.~(\ref{eq:recursion_final}), one can then write down the equations for the elements of the resolvent matrix on the cluster:
\[
\begin{aligned}
\frac{\mathfrak{g}_{2}^{p}}{{\rm det} \, {\cal G}_{p}}
& = -E_1 - z - \Gamma^2 \!\! \sum_{q_k \in \partial p}  \! g^{q_k \to p}_{1}  \, , \\
\frac{\mathfrak{g}_{1}^{p}}{{\rm det} \, {\cal G}_{p}}
& = -E_2 - z - \Gamma^2 \!\! \sum_{q_k \in \partial p}  \! g^{q_k \to p}_{2}  \, , \\
\frac{\mathfrak{g}_{12}^{p}}{{\rm det} \, {\cal G}_{p}}
& = -\Gamma + \Gamma^2 \!\! \sum_{q_k \in \partial p}  \! g^{q_k \to p}_{12}  \, , 
\end{aligned}
\]
where $\mathfrak{g}_{1}^{p}$ and $\mathfrak{g}_{2}^{p}$ are the diagonal elements on sites $i$ and $j$ of the cluster $p$ of the resolvent matrix, $\mathfrak{g}_{12}^{p}$ is the off-diagonal element, and ${\rm det} \, {\cal G}_{p} = \mathfrak{g}_{1}^{p} \mathfrak{g}_{2}^{p} - (\mathfrak{g}_{12}^{p})^2$
When 
$g_{12}^{p \to q_k}$ and $\mathfrak{g}_{12}^{p \to q_k}$ are set to zero, these equations give back the standard (cavity) recursion equations for the single-site Anderson model on the Bethe lattice.\cite{ATA,BiroliSemerjianTarzia}
Moreover, since the off-diagonal elements are proportional to $\Gamma$, in the limit of small transverse field one might expand these equations in powers of $\Gamma$ to obtain the systematic corrections to the zeroth-order equations due to the small loops: 
\[
\begin{aligned}
g_{1,0}^{p \to q_k} & = -E_1 - z - \Gamma^2 \!\!\!\! \sum_{q_l \in \partial p / q_k}  \!\!\!\! g^{q_l \to p}_{1,0}  \, , \\
g_{2,0}^{p \to q_k} & = -E_2 - z - \Gamma^2 \!\!\!\! \sum_{q_l \in \partial p / q_k}  \!\!\!\! g^{q_l \to p}_{2,0}  \, , \\
\delta g_{12,1}^{p \to q_k} & = - \Gamma g_{1,0}^{p \to q_k} g_{2,0}^{p \to q_k} + O (\Gamma^3) \, , \\
\delta g_{1,1}^{p \to q_k} & = \Gamma^2 \left ( g_{1,0}^{p \to q_k} \right)^2 g_{2,0}^{p \to q_k} + O (\Gamma^4) \, , \\
\delta g_{2,1}^{p \to q_k} & = \Gamma^2 \left ( g_{2,0}^{p \to q_k} \right)^2 g_{1,0}^{p \to q_k} + O (\Gamma^4)\, ,
\end{aligned}
\]
where $g_{1,0}^{p \to q_k}$ and $g_{2,0}^{p \to q_k}$ are the diagonal elements of the resolvent at the $0$-th order of the cluster expansion (i.e., within the standard single-site Bethe approximation), and $\delta g_{12,1}^{p \to q_k}$, $\delta g_{1,1}^{p \to q_k}$, and $\delta g_{2,1}^{p \to q_k}$ are the corrections for $n=1$ up to the lowest order in $\Gamma$.

\begin{figure}
\includegraphics[width=0.46\textwidth]{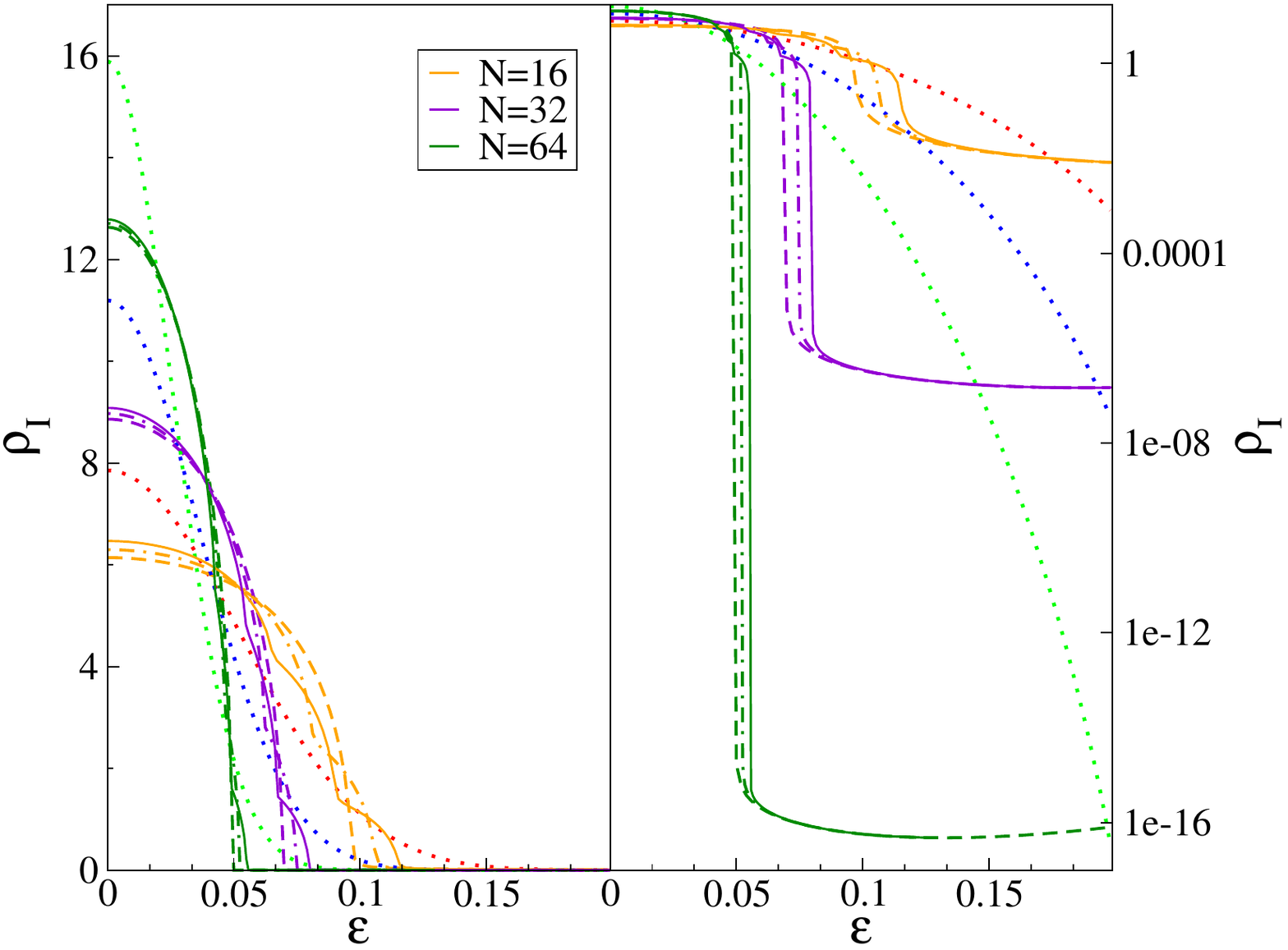}
\vspace{-0.2cm}
\caption{\label{fig:rhoI} Comparison between the exact DoS of the delocalizing interacting part of the QREM [i.e., $\Gamma$ times the adjacency matrix of the $N$ dimensional hypercube, Eq.~(\ref{eq:rhoIHC}), dotted red, blue, and green curves] and the spectra of the kinetic term of Eq.~(\ref{eq:H}) within the cluster approximation for $n=0$ [i.e., the standard single-site Bethe approximation, which yields ($\Gamma$ times) the adjacency matrix of a RRG of connectivity $N$, Eq.~(\ref{eq:rhoIRRG}), dashed orange, violet, and dark green curves], $n=1$ (dashed-dotted violet, and dark green curves), and $n=2$ (continuous violet, and dark green curves) for $\Gamma=0.2$ and three values of $N$. 
At each order of the cluster expansion the edge of the spectrum is shifted by $\Gamma$ to the right (to the leading order in $N$). The imaginary regulator $\eta$ is set to be $\eta = c \delta$, Eq.~(\ref{eq:eta}), with $\delta = 1/(2^N \rho(E))$.}
\end{figure}

One can proceed in a similar way for $n=2$ and obtain closed equations for the ten independent elements of the cavity resolvent matrices on each cluster\ in terms of the elements of the cavity resolvent matrices on the neighboring clusters. However the equations are much longer and we do not write them here explicitly. It is just worth to mention that in this case the off-diagonal elements of the resolvent matrix between pairs of sites of the cluster that are connected by an edge on the hypercube (e.g., $g_{12}^{p \to q_k}$, $g_{23}^{p \to q_k}$, $g_{34}^{p \to q_k}$, and $g_{14}^{p \to q_k}$, using the notation of fig.~\ref{fig:cluster}) are proportional to $\Gamma$. Conversely, the Green's functions between pairs of sites that are not connected by an edge on the hypercube (e.g., $g_{13}^{p \to q_k}$ and $g_{24}^{p \to q_k}$) are (as expected) proportional to $\Gamma^2$.

\subsection{Spectrum of the kinetic term}

It is instructive to study how the spectrum of the kinetic term of the Hamiltonian is modified by the cluster expansion.
To this aim we consider the pure limit (in absence of disorder) of the equations above for $n=1$. In the pure case ($E_i=E_j=0$) the hypercube is translationally invariant. One can thus look for a uniform solution of the equations in the form:
\[
\begin{aligned}
\frac{g}{g^2 - h^2}
& = - z - (N-2) \Gamma^2 g  \, , \\
\frac{g}{g^2 - h^2}
& = - \Gamma + (N-2) \Gamma^2 h  \, ,
\end{aligned}
\]
where $g = g_{1}^{p \to q_k} = g_{2}^{p \to q_k} $ and $h = g_{12}^{p \to q_k}$ for all $p,q$.
One can then introduce the variables $g_+ = g + h$ and $g_- = g - h$ in terms of which the equations above become:
\[
\begin{aligned}
\frac{1}{g_+} & = - z  + \Gamma - (N-2) \Gamma^2 g_+  \, , \\
\frac{1}{g_-} & = - z - \Gamma - (N-2) \Gamma^2 g_-  \, ,
\end{aligned}
\]
which coincide with the equations that one obtains for the standard single-site Anderson model on the Bethe lattice in the uniform limit, with energies shifted of $\pm \Gamma$. The DoS is thus modified accordingly. In particular the edges of the spectrum are shifted as well by $+\Gamma$ on the right edge and by $-\Gamma$ on the left edge.
One can indeed define $\mathfrak{g} = \mathfrak{g}_{1}^{p} = \mathfrak{g}_{2}^{p} $ and $\mathfrak{h} = \mathfrak{g}_{12}^{p}$ for all $p$, and introduce the variables $\mathfrak{g}_\pm = \mathfrak{g} \pm \mathfrak{h}$, 
which verify the following equations:
\[
\begin{aligned}
\frac{1}{\mathfrak{g}_\pm} & = - z  \pm \Gamma - (N-1) \Gamma^2 g_\pm  \, , 
\end{aligned}
\]
in terms of which the DoS can be obtained as $\rho_I^{(n=1)} = {\rm Im} ( \mathfrak{g}_+ + \mathfrak{g}_-)/(2 \pi)$.

For $n=2$ a similar treatment of the equations yields:
\[
\begin{aligned}
\frac{1}{g_\pm} & = - z  \pm 2 \Gamma - (N-3) \Gamma^2 g_\pm  \, , \\
\frac{1}{g_0} & = - z - (N-3) \Gamma^2 g_0  \, ,
\end{aligned}
\]
where $g_\pm = g + f \pm h$ and $g_0 = g - f$, where $g = g_{u}^{p \to q_k}$, $h = g_{12}^{p \to q_k} = g_{23}^{p \to q_k} = g_{34}^{p \to q_k} = g_{14}^{p \to q_k}$, and $f = g_{13}^{p \to q_k} = g_{24}^{p \to q_k}$ for all $u,p,q$.
Similarly one defines $\mathfrak{g} = \mathfrak{g}_{u}^{p}$, $\mathfrak{h} = \mathfrak{g}_{12}^{p} = \mathfrak{g}_{23}^{p} = \mathfrak{g}_{34}^{p} = \mathfrak{g}_{14}^{p}$, 
and $\mathfrak{f} = \mathfrak{g}_{13}^{p} = \mathfrak{g}_{24}^{p}$ for all $u,p$, and introduces the variables $\mathfrak{g}_\pm = \mathfrak{g} + \mathfrak{f} \pm \mathfrak{h}$ and $\mathfrak{g}_0 = \mathfrak{g} - \mathfrak{f}$, which verify the following equations:
\[
\begin{aligned}
\frac{1}{\mathfrak{g}_\pm} & = - z  \pm 2 \Gamma - (N-2) \Gamma^2 g_\pm  \, , \\
\frac{1}{\mathfrak{g}_0} & = - z - (N-2) \Gamma^2 g_0  \, ,
\end{aligned}
\]
in terms of which the DoS reads:
\[
\rho_I^{(n=2)} = {\rm Im} \left[ \frac{\mathfrak{g}_+ + \mathfrak{g}_- + 2 \mathfrak{g}_0}{4 \pi} \right] \, .
\]
A comparison between the exact spectrum of the kinetic term of the Hamiltonian~(\ref{eq:H}), i.e., the adjacency matrix of the $N$-dimensional hypercube~(\ref{eq:rhoIHC}), the DoS resulting from the cluster approximation for $n=0$, i.e., the adjacency matrix of a RRG of connectivity $N$, Eq.~(\ref{eq:rhoIRRG}), $n=1$, and $n=2$ is shown in fig.~\ref{fig:rhoI} for $\Gamma=0.2$ and three values of $N$.
The imaginary regulator is set to the value used to solve the recursion equations within the C-BP approximation in presence of the disordered  many-body  on site energies, Eq.~(\ref{eq:eta}).
At the order $n$ of the cluster expansion the (right) edge of the spectrum is shifted by $+n\Gamma$ ($-n\Gamma$) to the right (left) to the leading order in $N$.

\section{Solution of the recursion equations and comparison with exact diagonalization} \label{app:cavity}

As explained in the main text, the cluster approximation 
allows us to derive a system of closed equations, Eqs.~(\ref{eq:recursion}) and~(\ref{eq:recursion_final}), for the diagonal elements of the resolvent matrix of~(\ref{eq:H}).
A first, and crucial, question that we want to address here is to what extent this approximation provides a good qualitative and quantitative description of the spectral statistics of the QREM.
To this aim in this section we consider samples of moderate size and compare the probability distributions of the LDoS computed from the numerical solution of the recursion relations~(\ref{eq:recursion}) and~(\ref{eq:recursion_final}) with those obtained from EDs of the QREM.

There are essentially two ways, that we detail below, to solve the recursion equations for the Green's function and obtain information on the spectral statistics at finite $N$. 

\vspace{0.2cm}
\noindent {\bf 1) C-BP on the hypercube.} The most accurate strategy, to which we will refer to as ``Cluster Belief Propagation'' (C-BP) algorithm (see Ref.~[\onlinecite{BiroliTarzia-Review}] for a detailed explanation of this approach for the usual tight-binding Anderson model on the Bethe lattice), is to solve directly Eqs.~(\ref{eq:recursion}) and~(\ref{eq:recursion_final}) on random realizations of the hypercube of $2^N$ sites (i.e., $N$ spins). 
In practice, one proceeds as follows: 
\begin{itemize}
\item One first generates a random instance of the hypercube drawing the $2^N$ on-site energies from the distribution~(\ref{eq:PE});
\item One finds a partition of the hypercube in $2^{N-n}$ clusters of $2^n$ sites each (note that the choice of the partition is {\it not} unique);
\item Then one finds the fixed point of Eqs.~(\ref{eq:recursion}), which constitute a system of $(s+1) (N-n) 2^{N-1}$ coupled equation for the $s(s+1)/2$ independent elements of the cavity Green's functions on each cluster (this can be done iteratively with arbitrary precision in a time which scales linearly with $(s+1) (N-n) 2^{N-1}$);
\item Using Eqs.~(\ref{eq:recursion_final}) one obtains the $s(s+1)/2$ independent elements of the resolvent matrix on each cluster of that specific instance (and for that specific choice of the partition of the hypercube in clusters).  
\item One then repeat this procedure several times to average over different realizations of the on-site disorder (and over different choices of the cluster partitioning). 
\end{itemize}
As discussed above (see also~[\onlinecite{BiroliTarzia-Review}]), in order for the recursive equations to converge to the physical fixed point, the broadening $\eta$ must be larger than the mean level spacing. Hence, in order to implement the $\eta \to 0^+$ limit correctly, for any given choice of the parameters $\Gamma$, $\varepsilon$ and $N$, the imaginary regulator is self-consistently set to be a constant of order $1$ times the mean level spacing:
\begin{equation} \label{eq:eta}
\eta = \frac{c}{2^N \rho_\eta (\varepsilon)} = \frac{c \pi}{\sum_{i=1}^\vol {\rm Im} {\cal G}_i (N \varepsilon + i \eta)} \, .
\end{equation}
As shown in~[\onlinecite{BiroliTarzia-Review}], for large enough system sizes and for $c$ large enough, $\rho_\eta$ converges to its asymptotic value obtained in the limit $\vol \to \infty$ and $\eta \to 0^+$. We will take $c=64$ throughout.\footnote{We have checked that varying $c$ from $16$ to $128$ do not modify the results.} 
Within the C-BP approach one can study hypercubes of sizes up to $N=26$, which are considerably larger than the ones accessible via the most efficient ED algorithms.

\vspace{0.2cm}
\noindent {\bf 2) C-PD algorithm on the RRG.} In order to access even larger system sizes, one can adopt another strategy, to which we will refer hereafter as ``Cluster Population Dynamics'' (C-PD) algorithm,\cite{MezardParisi2001PopDyn} which consists in interpreting
the recursion relations for the Green's functions as equations for their probability distributions once the average over the disorder is taken. 
In fact, since $G_{p \to q}$ and ${\cal G}_{p}$ are random matrices, one can assume that averaging over the on-site random energies 
leads to 
functional equations on their probability distribution $Q (G)$ and ${\cal Q}({\cal G})$. From Eq.~(\ref{eq:recursion}) we naturally get:
\begin{equation} \label{eq:PGcav}
\begin{split}
Q (G) &= \! \int \! \prod_{u=1}^{2^n} \textrm{d} P(E_u) 
	\prod_{q=1}^{N-n-1} \textrm{d} Q (G_q) \\
& \,\,\,\,\,\,\,\,\,\, \times 
\delta \! \left ( \! G^{-1} \! + {\cal H}_c + z {\cal I}_s  + \Gamma^2 \sum_{q=1}^{N-n-1} G_q \! \right) \, ,
\end{split}
\end{equation}
where $P(E)$ is given by Eq.~(\ref{eq:PE}), ${\cal H}_c$ is the Hamiltonian~(\ref{eq:H}) on the sites of the cluster (see, e.g., Appendix~\ref{app:clusters} for the explicit expression of ${\cal H}_c$ for $n=1$), which contains the $s$ diagonal random energies $E_1, \ldots, E_s$, and ${\cal I}_s$ is the $s \times s$ identity matrix. (The notation $\textrm{d} Q (G_q)$ is just a shortcut for the integration over the $s(s+1)/2$ independent elements of $G_q$.)
Once the fixed point of this equation is obtained, using Eq.~(\ref{eq:recursion_final})
one can find an equation for the probability distribution of the 
elements of the resolvent:
\begin{equation} \label{eq:PG}
\begin{split}
	{\cal Q} ({\cal G}) &= \! \int \! \prod_{u=1}^{2^n} \textrm{d} P(E_u) 
	\prod_{i=q}^{N-n} \textrm{d} Q (G_q) \\
& \,\,\,\,\,\,\,\,\,\, \times 
        \delta \! \left ( \! {\cal G}^{-1} \! + {\cal H}_c + z {\cal I}_s + \Gamma^2 \sum_{q=1}^{N-n} G_q \! \right) \, .
\end{split}
\end{equation}
As before, $z = N \varepsilon + i \eta$ and the imaginary regulator is self-consistently set to be $c$ times the mean level spacing:
$\eta = 2^{-N} c \pi / \langle {\rm Im} {\cal G} \rangle$, where the average is performed over the distribution ${\cal Q} ({\cal G})$.
This set of functional equations can be solved numerically with an arbitrary degree of precision
using a population dynamics algorithm~\cite{BiroliSemerjianTarzia,BiroliTarzia-Review,Altshuler2016,MezardParisi2001PopDyn}. Hereafter we will show results obtained using populations of $M$ fields going from $M=2^{25}$ to $M=2^{27}$ (and for $n=2$).
Note that, differently from the C-BP approach, within the C-PD approximation the specific structure of the hypercube is completely lost for distances larger than the size of the clusters (apart from the local connectivity of each cluster equal to $N-n$).

\vspace{0.2cm}
On the other hand, from ED\footnote{In fact, the diagonal elements of the resolvent can be obtained by matrix inversion, which is slightly faster than ED.} we can easily obtain the matrix elements ${\cal G}_i$ for a given instance of the QREM in terms of the eigenvalues $E_\alpha$ and the eigenvectors $\vert \alpha \rangle$ of~(\ref{eq:HQREM}) as:
\begin{equation} \label{eq:greenED}
{\cal G}_i (N \varepsilon + i \eta) = \sum_{\alpha=1}^{2^N} \vert \langle \alpha \vert i \rangle \vert^2 \frac{E_\alpha - N \varepsilon + i \eta}{(E_\alpha - N \varepsilon)^2 + \eta^2} \, .
\end{equation}
For each choice of the parameters $\Gamma$, $N$, and $\varepsilon$, the imaginary regulator is set to the same value as the one used to solve the self-consistent recursion relations,  Eq.~(\ref{eq:eta}).

\begin{figure}
\includegraphics[width=0.49\textwidth]{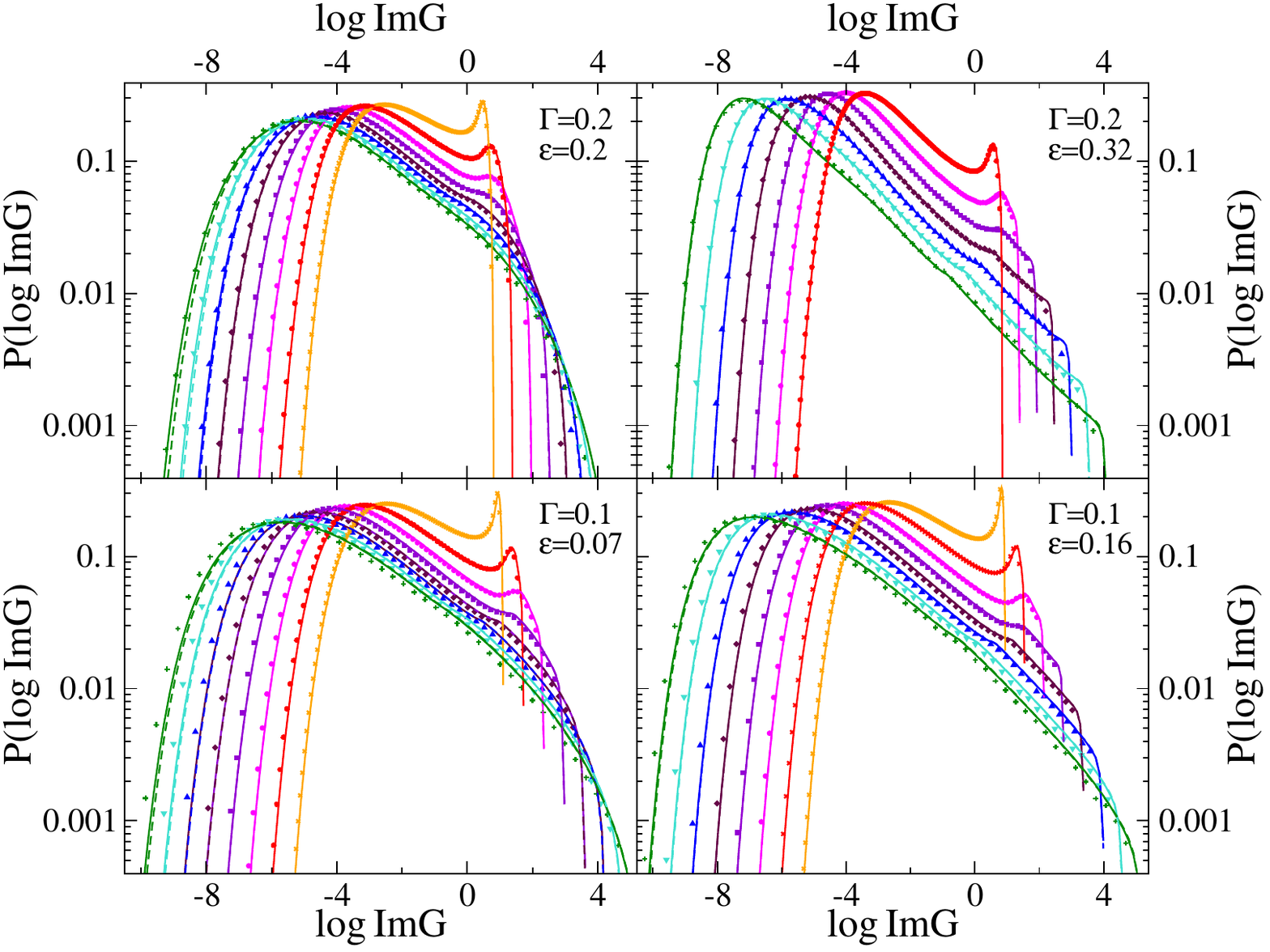}
\caption{\label{fig:PLIG} Probability distributions $P(\ln {\rm Im}G)$ obtained from ED (filled symbols), C-BP ($n=2$, continuous curves), and C-PD ($n=2$, dashed curves) for $N=8$ (orange), $10$ (red), $11$ (violet), $12$ (maroon), $13$ (blue), $14$ (turquoise), and $15$ (green), and for $\Gamma=0.2$ and $\varepsilon=0.2$ (top left panel), $\Gamma=0.2$ and $\varepsilon=0.32$ (top right panel), $\Gamma=0.1$ and $\varepsilon=0.07$ (bottom left panel), $\Gamma=0.1$ and $\varepsilon=0.16$ (bottom right panel).}
\end{figure}

In fig.~\ref{fig:PLIG} we focus on the probability distribution of the imaginary part of the Green's functions, and plot ${\cal Q} (\ln {\rm Im}{\cal G})$ for $\Gamma=0.1$ and $\Gamma=0.2$, for $N$ ranging from $10$ to $15$, and for two values of $\varepsilon$ which are supposed to be on the ergodic side of the MBL transition and close to the mobility edge respectively.\cite{Laumann2014,Baldwin2016}  
In all cases, we observe a good agreement between the probability distributions found  
from EDs and Eq.~(\ref{eq:greenED}), and their C-BP counterpart, found from the numerical solution of the recursion relations~(\ref{eq:recursion}) and~(\ref{eq:recursion_final}) on the hypercube for $n=2$. 
In the figure we also plot the distributions ${\cal Q}(\ln {\rm Im} {\cal G})$ obtained from the C-PD algorithm, Eqs.~(\ref{eq:PGcav}) and~(\ref{eq:PG}), which presents very small deviations from the C-BP results only in the very far tails of the distributions at small ${\rm Im} {\cal G}$, and only visible for some values of $\Gamma$, $\varepsilon$, and $N$.

\section{Analytic computation of the localization threshold of the auxiliary Anderson models in the large-\texorpdfstring{$N$}{N} limit} \label{app:analytic}

In this appendix we discuss the analytical computation of the localization threshold(s) of the family of auxiliary Anderson tight-binding models described by the Hamiltonian~(\ref{eq:H}), when the thermodynamic limit, $\vol \to \infty$, is taken from the start, while keeping $N$ fixed.
For simplicity, we will only consider the simplest setting $n=0$, i.e., the standard single-site Bethe approximation in which the hypercube is approximated by a tree-like structure of connectivity $N$.

\subsection{Probability distribution of the real part of the self-energy} \label{app:real}

\begin{figure}
\includegraphics[width=0.49\textwidth]{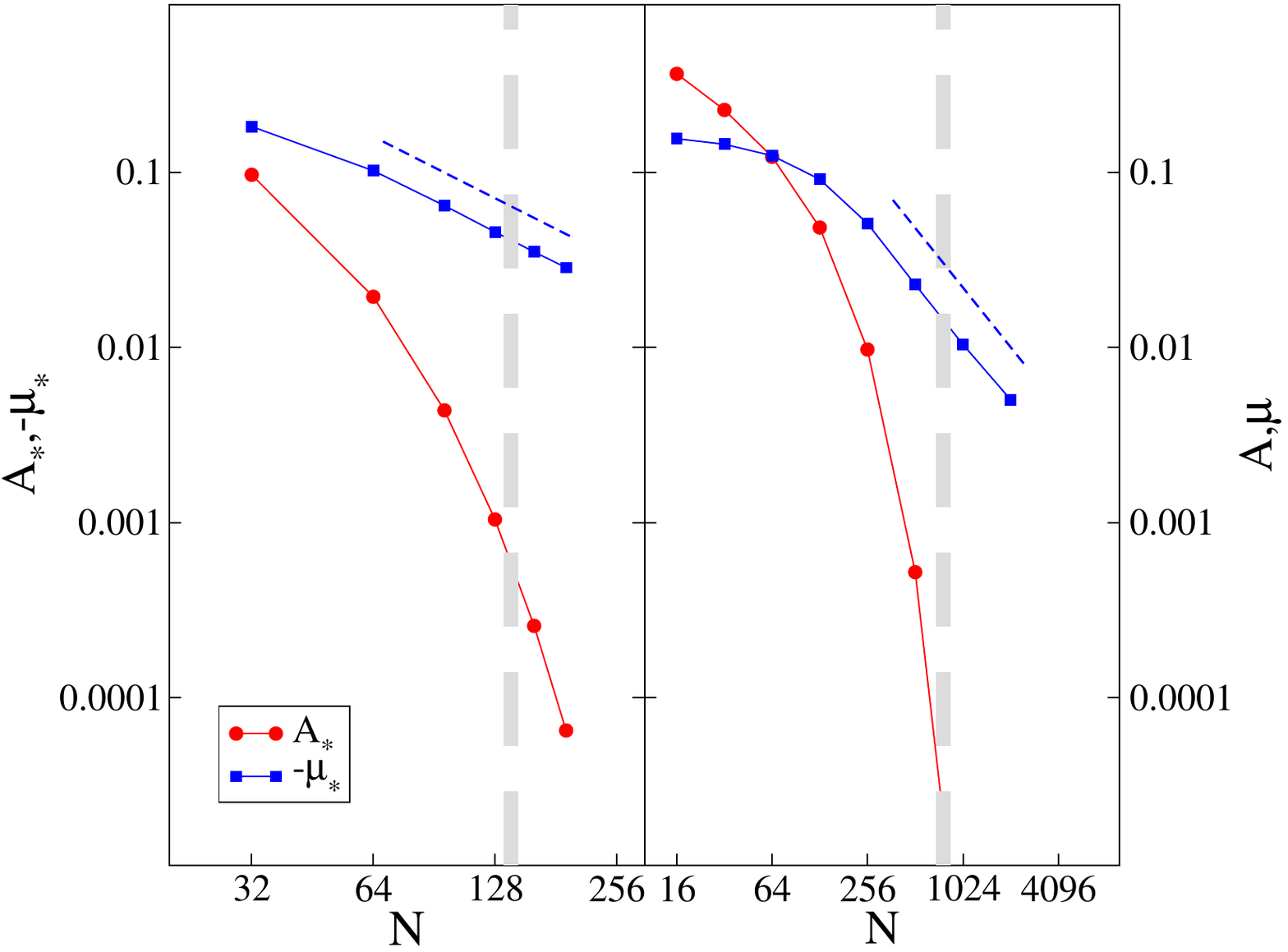}
\vspace{-0.2cm}
\caption{\label{fig:REALPART} Solutions $A_\star$ (red) and (minus) $\mu_\star$ (blue) of Eqs.~(\ref{eq:A-mu}) for $\Gamma=0.2$ and $\varepsilon=0.2$ (left panel) and $\varepsilon=0.1$ (right panel) as a function of $N$. The blue dashed lines correspond to $\mu_\star \sim - 1/N$. The vertical thick gray dashed lines indicate the localization threshold where the Lyapunov exponent vanishes for these particular values of $\Gamma$ and $\varepsilon$ (see fig.~\ref{fig:lyap}).}
\end{figure}

The fisrt step is to realize that the recursion relations for the real part of the self-energies in the linearized regime is independent on the imaginary part, and can be solved as explained below. 
It is useful to introduce the variables $X_{i \to j}$ (i.e., the real part of the diagonal elements of the resolvent matrix in the linearized regime) defined as:
\begin{equation} \label{eq:X}
X_{i \to j} = - \frac{1}{E_{i} + N \varepsilon + S_{i \to j}} = G^R_{i \to j} \, .
\end{equation}
In terms of these variables at the $0$-th order of the cluster expansion Eqs.~(\ref{eq:SigmaLinR}) and~(\ref{eq:SigmaLin}) become:
\begin{equation} \label{eq:SigmaX}
\begin{aligned}
S_{i \to j} & = \Gamma^2 \!\! \sum_{j^\prime \in \partial i / j} 
X_{j^\prime \to i} \, , \\
\Delta_{i \to j} & = \Gamma^2 \!\! \sum_{j^\prime \in \partial i / j}  X_{j^\prime \to i}^2 \, 
\Delta_{j^\prime \to i} \, .
\end{aligned}
\end{equation}
Hence, the probability distribution of the real part of the self-energy $R_S(S)$ can be obtained in terms of the probability distribution $R_X(X)$:
\[
\begin{aligned}
R_S (S) & = \int \prod_{i=1}^{N-1} \textrm{d} X_i \, R_X (X_i) \, \delta \left( S - \Gamma^2 
\sum_i X_i \right) \\
& = \int \frac{\textrm{d}k}{2 \pi} e^{i k S}
\left[ \prod_i \frac{\textrm{d} x_i \textrm{d} k_i}{2 \pi} \, e^{i ( k_i - \Gamma^2 k) x_i} \, 
\hat{R}_X (k_i) \right] \, ,
\end{aligned}
\]
where $\hat{R}_X (k)$ is the characteristic function of $R_X (X)$. Assuming that at small $k$ it behaves as the characteristic function of a Cauchy distribution,
\begin{equation} \label{eq:characteristicX}
\hat{R}_X (k) \simeq 1 - A |k| - i k \mu \, ,
\end{equation}
implies that in the large $N$ limit also $R_S(S)$ is given by a Cauchy distribution:
\begin{equation} \label{eq:PS}
\begin{aligned}
\hat{R}_S (k) &= \left[ \hat{R}_X ( \Gamma^2 k ) \right]^{N-1} \simeq e^{-(N-1) A \Gamma^2 |k| - i (N-1) \Gamma^2 \mu k} \, , \\
R_S(S) &= \frac{A_S}{\pi \left[ \left( S - \mu_S \right)^2 + A_S^2 \right]} \, ,
\end{aligned}
\end{equation}
with
\begin{equation} \label{eq:As-mus}
A_S = N \Gamma^2 A \, , \qquad \textrm{and} \qquad \mu_S = N \Gamma^2 \mu \, .
\end{equation}
(Throughout we will consider the large $N$ limit $N-1 \approx N$.)
On the other hand, from Eq.~(\ref{eq:X}) we have that:
\begin{equation} \label{eq:PXself}
\begin{aligned}
R_X (X) & = \int \textrm{d} E \, \textrm{d} S \, P (E) \, R_S(S) \, \delta \left (X + \frac{1}{E + N \varepsilon + S} \right) \\
& =  \frac{1}{|X|^2} \int \textrm{d} E \, P(E) \, R_S \left(- E - N \varepsilon - \frac{1}{X} \right ) \, .
\end{aligned}
\end{equation}
After some simple algebra we obtain that:
\begin{equation}
R_S \left(- E - N \varepsilon - \frac{1}{X} \right ) 
= \frac{c X^2}{\pi \left[ \left(X - X_0 \right)^2 + c^2 \right]} \, ,
\end{equation}
with
\[
\begin{aligned}
c & = \frac{A_S}{\left(E + N \varepsilon + \mu_S \right)^2 + A_S^2} \, , \\
X_0 &= - \frac{E + N \varepsilon + \mu_S}{\left(E + N \varepsilon + \mu_S \right)^2 + A_S^2} \, .
\end{aligned}
\]
As a result, from the second line of Eq.~(\ref{eq:PXself}) and from the relations above, we get:
\begin{equation}
R_X (X) = \int \textrm{d} E \, P (E) \, \frac{c}{\pi \left[ \left(X - X_0 \right)^2 + c^2 \right]} \, .
\end{equation}
We can now finally compute self-consistently the characteristic function of $R_X(X)$ by expanding the equation above up to first order in $k$:
\[
\begin{aligned}
\hat{R}_X (k) & = \int \textrm{d} E \, P (E) \, e^{-c |k| - i k X_0} \\
& \simeq 1 - \int \textrm{d} E \, P (E) \left(c |k| + i k X_0 \right ) \, .
\end{aligned}
\]
From Eqs.~(\ref{eq:characteristicX}), (\ref{eq:As-mus}), and the last equation we can thus obtain two self-consistent relations for the coefficients $A$ and $\mu$:
\begin{equation} \label{eq:A-mu}
\begin{aligned}
A_{\star} & = \int \textrm{d} E \, P (E_i) \, \frac{N \Gamma^2 A_\star}{\left(E + N \varepsilon + N \Gamma^2 \mu_\star \right)^2 + 
(N \Gamma^2 A_\star)^2} \, , \\
\mu_\star & = - \int \textrm{d} E \, P (E) \, \frac{E + N \varepsilon + N \Gamma^2 \mu_\star}
{\left(E + N \varepsilon + N \Gamma^2 \mu_\star \right)^2  + (N \Gamma^2 A_\star)^2} \, .
\end{aligned}
\end{equation}
These equations can be easily solved numerically.
In fig.~\ref{fig:REALPART} we show the solutions $A_\star$ and (minus) $\mu_\star$ of Eqs.~(\ref{eq:A-mu}) for $\Gamma=0.2$ and $\varepsilon=0.2$ (left panel) and $\varepsilon=0.1$ (right panel) as a function of $N$. While $A_\star$ decay very fast (exponentially) with $N$, $\mu_\star$ decreases much more slowly, roughly as $1/N$ (blue dashed lines). The vertical thick gray dashed lines indicate the localization threshold where the Lyapunov exponent vanishes for these particular values of $\Gamma$ and $\varepsilon$ (see fig.~\ref{fig:lyap}). 

\begin{figure}
\includegraphics[width=0.49\textwidth]{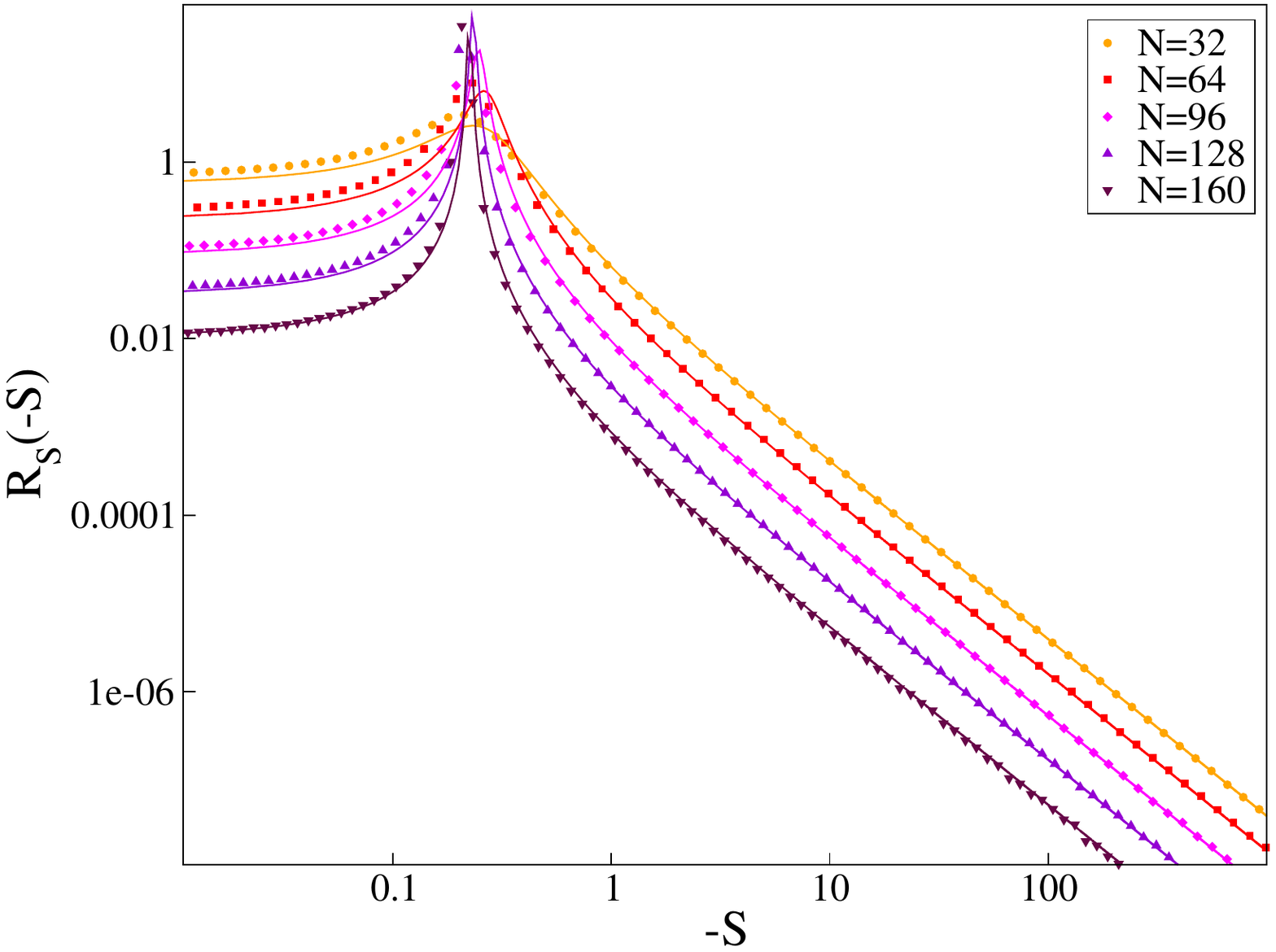}
\vspace{-0.2cm}
\caption{\label{fig:RS} Probability distributions $R_S(-S)$ for $\Gamma=0.2$, $\varepsilon=0.2$, and several values of $N$ across the localization transition of the auxiliary models. Filled symbols are obtained as the numerical solution of the linearized recursion relations for the self energy with the C-PD algorithm for $n=0$, while continuous lines correspond to the analytic prediction of Eqs.~(\ref{eq:PS}),~(\ref{eq:As-mus}), and~(\ref{eq:A-mu}).}
\end{figure}

In fig.~\ref{fig:RS} we plot the probability distribution of the real part of the self-energy, $R_S(S)$, for $\Gamma=0.2$, $\varepsilon=0.2$, and several values of $N$ across the localization threshold. We focus on the negative real axis since the peak of the distribution is located in $S= \mu_S$ which turns out to be negative. Filled symbols correspond to the numerical solution found using the C-PD algorithm (for $n=0$) of the linearized recursion equations for the self-energy~(\ref{eq:SigmaLin}), while the continuous lines correspond to the analytic prediction~(\ref{eq:PS}), with $A_S$ and $\mu_S$ given by Eqs.~(\ref{eq:As-mus}) and~(\ref{eq:A-mu}). The agreement between the numerical results and the analytic solution is excellent, and it improves for large $N$.

\subsection{Computation of the Lyapunov exponent}

Once the probability distribution of the real part of the self-energy has been obtained, we can focus on the integral equation for the joint distributions of the real and the imaginary part (for $n=0$):
\begin{equation}
\begin{aligned}
Q(S,\Delta) & = \int \prod_{i=1}^{N-1} \left[ \textrm{d} E_i \, P(E) \, \textrm{d} S_i \, {\rm d} \Delta_i \, Q (S_i,\Delta_i) \right] \\
& \qquad \times \delta \left( S + \Gamma^2 \sum_i \frac{1}{E_i + N \varepsilon + S_i} \right) \\
& \qquad \times \delta \left( \Delta - \Gamma^2 \sum_i \frac{\Delta_i}{(E_i + N \varepsilon + S_i)^2} \right) \, .
\end{aligned}
\end{equation}
We replace the $\delta$-functions by their integral representation in the Fourier space and also
write $Q(S_i,\Delta_i)$ as the inverse Fourier transform of $\hat{Q}_2 (S_i,k_i)$ with respect to the second argument, defined as:
\begin{equation} \label{fourier}
\hat{Q}_2 (S,k) = \int_{-\infty}^{+\infty} {\rm d} \Delta \, e^{-i k \Delta} \, Q(S,\Delta) \, ,
\end{equation}
yielding:
\[
\begin{aligned}
\hat{Q} (k_1,k_2) & = \int \prod_{i=1} \bigg[ \textrm{d} E_i \, P(E_i) \, \frac{\textrm{d} S_i \, {\rm d} \Delta_i \, {\rm d}k_i}{2 \pi}
\, \hat{Q}_2 (S_i,k_i) \\
& \,\,\, \times e^{i k_1 \Gamma^2 / (E_i + N \varepsilon + S_i)} \, e^{\Delta_i[k_i - k_2 \Gamma^2/(E_i + N \varepsilon + S_i)^2]} \bigg ]\, .
\end{aligned}
\]
We can now perform the integration over $\di \Delta_i$, which gives
$2 \pi \delta ( k_i - k_2 \Gamma^2 / (E_i + N \varepsilon + S_i)^2 )$, and then integrate over $k_i$:
\begin{equation} \label{Qk1k2}
\begin{aligned}
\hat{Q} (k_1,k_2) & = \bigg[ \int \textrm{d} E \, P(E) \, \textrm{d} S \, \hat{Q}_2 \left( S,\frac{k_2 \Gamma^2}{(E + N \varepsilon + S)^2}
\right) \\
& \qquad \qquad \times e^{i k_1 \Gamma^2 / (E + N \varepsilon + S)} \bigg]^{N-1} \, .
\end{aligned}
\end{equation}
Similarly to Ref.~[\onlinecite{ATA}], we assume that in the localized phase the following asymptotic form
of $Q(S,\Delta)$ holds for large enough $\Delta$:
\begin{equation} \label{asympt}
\begin{aligned}
Q(S,\Delta) & \simeq \frac{A(S)}{\Delta^{1+\beta}} \qquad {\rm for~} \Delta \to \infty\, , \\
\hat{Q}_2 (S,k) & \simeq \hat{Q}_2 (S,0) - \alpha |k|^\beta A(S) \qquad {\rm for~} k \to 0 \, ,
\end{aligned}
\end{equation}
where $\hat{Q}_2 (S,0)$ is by definition the marginal of $Q(S,\Delta)$
once we integrate over $\Delta$, i.e., $Q(S,\Delta) = R_S (S)$.
Plugging the asymptotic form~(\ref{asympt}) into both sides of Eq.~(\ref{Qk1k2}) we obtain:
\begin{equation} \label{eq:expansion}
\begin{aligned}
& \hat{R}_S(k_1) - \alpha |k_2|^\beta \hat{A} (k_1) \simeq \bigg[ \int \textrm{d} E \, P(E) \, \textrm{d} S \, \bigg( R_S(S) \\
& \qquad \,\,\, - \alpha \left \vert
\frac{k_2 \Gamma^2}{(E + N \varepsilon + S)^2} \right \vert^\beta A(S) \bigg) e^{i k_1 \Gamma^2 / (E + N \varepsilon + S)} 
\bigg]^{N-1}
\end{aligned}
\end{equation}
We can now expand the right-hand side of Eq.~\eqref{eq:expansion} in powers of 
$k_2$ up to the order $|k_2|^\beta$ and define:
\[
\begin{aligned}
I_1 & = \int \textrm{d} E \, P(E) \, \textrm{d} S \, R_S(S) e^{i k_1 \Gamma^2 / (E + N \varepsilon + S)} \, , \\
I_2 & = \int \textrm{d} E \, P(E) \, \textrm{d} S \, 
\frac{\Gamma^{2 \beta}}{\vert E + N \varepsilon + S \vert^{2 \beta}} A(S) \, e^{i k_1 \Gamma^2/ (E + N \varepsilon + S)} \, .
\end{aligned}
\]
The r.h.s. of Eq.~(\ref{eq:expansion}) is given by $(I_1 - \alpha |k_2|^\beta I_2)^{N-1} \simeq I_1^{N-1} [ 1 - \alpha (N-1) |k_2|^\beta
I_2 / I_1 ] = I_1^{N-1} - \alpha (N-1) |k_2|^\beta I_1^{N-2} I_2$.
From Eq.~(\ref{eq:SigmaLin}) we have that by definition $I_1^{N-1} = \hat{R}_S$.
Hence, in the large-$N$ limit, $I_1^{N-2} \simeq I_1^{N-1} = \hat{R}_S (k_1)$, we get:
\[
\begin{aligned}
\hat{A} (k_1) & \simeq N \Gamma^{2 \beta} \hat{R}_S (k_1) \int \textrm{d} E \, P(E) \, \textrm{d} S \\
 & \qquad \qquad \qquad \times A(S) \, \frac{e^{i k_1 \Gamma^2 / (E + N \varepsilon + S)}}{\vert E + N \varepsilon + S \vert^{2 \beta}} \, .
\end{aligned}
\]
Changing variable to $w = E + N \varepsilon + S$, replacing $A(S)$ by the inverse Fourier transform of $\hat{A}(k)$ and integrating
over ${\rm d} E$ we obtain:
\begin{equation}
\begin{aligned}
\label{eq:integral}
\hat{A} (k_1) & \simeq N \Gamma^{2 \beta} \hat{R}_S (k_1) \int \frac{{\rm d} w \, {\rm d} k}{2 \pi} \, e^{- \frac{N k^2}{4} - i k N \varepsilon} \\ & \qquad \qquad \qquad \times  \hat{A} (k) \frac{e^{i k w + i k_1 \Gamma^2/w}}{| w |^{2 \beta}} \, .
\end{aligned}
\end{equation}
For a given choice of the parameters $\Gamma$, $\varepsilon$, and $N$, the localization threshold at the $0$-th order of the cluster expansion is thus given by the value of the energy $\varepsilon_{\rm loc}(\Gamma,\varepsilon,N)$ such that the largest eigenvalue $\lambda_\beta$ of the integral operator defined by the equation above, becomes equal to one. As first noticed in Ref.~[\onlinecite{ATA}] (see also Refs.~[\onlinecite{Bapst2014}],~[\onlinecite{Parisi2019AndersonBethe}], and~[\onlinecite{AltshulerPrigodin1989}]) the kernel of the integral operator is symmetric around $\beta=1/2$ 
under the transformation $\beta \to 1 - \beta$, which implies that $\lambda = 1$ if and only if $\beta = 1/2$.
Since this is the value of interest for the transition, hereafter we will focus on the case $\beta=1/2$ only. The integral over ${\rm d} w$ can then be performed in terms of modified Bessel functions:
\[
\int {\rm d} w \, \frac{e^{i k w + i k_1 \Gamma^2/w}}{| w |} = - 2 \pi Y_0 \left (2 \Gamma \sqrt{k k_1} \right) \, .
\]
Following Ref.~[\onlinecite{Bapst2014}] we now assume that in the large connectivity limit the eigenvector of the integral operator defined 
by Eq.~(\ref{eq:integral}) for $\beta=1/2$ is very well approximated by $\hat{R} (k)$.
Assuming that the localization transition occurs on such energy scales (apart from logarithmic corrections)
\begin{equation} \label{eq:epstilde}
\varepsilon = \frac{\tilde{\varepsilon}}{\sqrt{N}} \, ,
\end{equation}
with $\tilde{\varepsilon}$ of $O(1)$, 
the equation for the mobility edge becomes:
\[
\begin{aligned}
1 &= N \Gamma \int \frac{{\rm d} k}{2 \pi} e^{- \frac{N k^2}{4} - N \Gamma^2 A_\star |k| - i k ( \sqrt{N} \tilde{\varepsilon} + N \Gamma^2 \mu_\star )} \\
&\qquad \qquad \qquad \qquad \times \left [ - 2 \pi Y_0 \left ( 2 \Gamma \sqrt{k k_1} \right ) \right] \, ,
\end{aligned}
\]
where we have used Eqs.~(\ref{eq:As-mus}).
$A_\star$ is exponentially small in $N$ and can be neglected, while $N \mu_\star$ is of order $1$ at the transition (see fig.~\ref{fig:REALPART}), and gives a correction 
of order $1/\sqrt{N}$ 
to $\tilde{\varepsilon}$ as $\tilde{\varepsilon}^\prime = \tilde{\varepsilon} + \Gamma^2 \sqrt{N} \mu_\star$. 
Since the integral over $k$ is cut-off on a scale $1/\sqrt{N}$ we can then expand the Bessel function keeping only  the leading logarithmic divergence at small $k$, $Y_0 (x) \approx 2 \ln(x) / \pi$.
In the $N \to \infty$ limit we can then change variable to $\tilde{k} = \sqrt{N} k$, yielding:
\[
1 = \frac{\sqrt{N} \Gamma}{\sqrt{\pi}} \, e^{- (\tilde{\varepsilon}^\prime)^2} \left ( \ln N - 4 \ln \Gamma + O(1) \right ) \, .
\]
Putting everything together we finally obtain the equation for the mobility edge~(\ref{eq:mobility}) given in the main text.

\subsection{The Forward-Scattering Approximation}

The FSA consists in neglecting the real part of the self-energy in the denominators of Eqs.~(\ref{eq:SigmaLinR}) and~(\ref{eq:SigmaLin}).
From Eq.~(\ref{eq:PXself}), one can then compute the probability distribution $R_X (X)$ for $n=0$:
\[
R_X(X) = \frac{1}{|X|^2} \, \frac{e^{-\frac{1}{N} \left( \frac{1}{X} - N \varepsilon \right)^2}}{\sqrt{\pi N}} \, ,
\]
which does not verify exactly the asymptotic form~(\ref{eq:characteristicX}) for its behavior at large $X$ in presence of the real part $S$.
This implies that, differently from what happens for the large connectivity limit of the usual Anderson tight-binding model on the Bethe lattice,  the distribution $R_S (S)$ found within the FSA does not coincides exactly  with the distribution obtained in Sec.~\ref{app:real} in presence of the real parts in the denominators.

Following the steps of the calculation detailed above for the largest eigenvalue of the linearized recursion relations, one can compute the distribution of the imaginary part of the self-energy as:
\[
\begin{aligned}
Q(\Delta) & = \int \prod_{i=1}^{N-1} \left[ \textrm{d} E_i \, P(E) \, \Delta_i \, Q (\Delta_i) \right] \\
& \qquad \qquad \times \delta \left( \Delta - \Gamma^2 \sum_i \frac{\Delta_i}{(E_i + N \varepsilon)^2} \right) \, ,
\end{aligned}
\]
which gives the following self-consistent equation for its Fourier transform:
\[
\hat{Q} (k) = \left[ \int {\rm d} E \, P(E) \, \hat{Q} \left( \frac{k \Gamma^2}{(E + N \varepsilon)^2} \right)\right]^{N-1} \, .
\]
Assuming, as before, the asymptotic form $\hat{Q} (k) \approx 1 - \alpha \vert  k \vert^\beta$, and expanding the last equation at large $N$, one gets the equation for the localization threshold within the FSA:
\begin{equation} \label{eq:FSA}
1 \approx N \Gamma^{2 \beta} \int {\rm d} E \, P(E) \, \frac{1}{\vert E + N \varepsilon \vert^{2 \beta}} \, .
\end{equation}
The symmetry $\lambda(\beta) = \lambda(1-\beta)$ is now lost. Hence the transition point is not achieved at $\beta = 1/2$, but rather at a given point $\beta_\star \in [0,1/2]$ which depend on the other parameters of the auxiliary models. 
Eq.~(\ref{eq:FSA}) can be easily solved numerically for any choice of $\Gamma$, $\varepsilon$, $N$, and $\beta$, and gives the localization
threshold $\varepsilon_{\rm loc}^{\rm FSA} = \Gamma$, with $\beta_\star \to 1/2$ for $N \to \infty$.
Indeed, since the random energy $E$ is typically of order $\sqrt{N}$, if $\varepsilon$ is of order $1$, one can expand the denominator in powers of $E/(N \varepsilon)$ yielding:
\[
1 \approx N^{1-2 \beta} \left ( \frac{\Gamma}{\varepsilon} \right)^{2 \beta} \left( 1 + \frac{\beta (2 \beta - 1)}{2 N \varepsilon^2} + \ldots \right)\, ,
\]
which, in the $N \to \infty$ limit and $\beta \to 1/2$, gives $\varepsilon_{\rm loc}^{\rm FSA} = \Gamma$.

\bibliographystyle{apsrev4-2}
\bibliography{QREM.bib}

\end{document}